\PassOptionsToPackage{hyphens}{url}

\documentclass[a4paper,fleqn]{cas-sc}
\usepackage[numbers,sort&compress]{natbib}
\usepackage{placeins}
\usepackage{capt-of}
\IfFileExists{xurl.sty}{\usepackage{xurl}}{}

\hypersetup{hypertexnames=false}

\ExplSyntaxOn

\cs_set:Npn \__first_footerline: {\mbox{}}
\ExplSyntaxOff

\begin{document}

\shorttitle{Trajectory-unsupervised neural solvers for molecular dynamics}
\title[mode=title]{Towards trajectory-unsupervised physics-informed neural solvers for molecular dynamics}

\author[1]{Petros Triantafyllos}

\author[2]{Panagiotis Krokidas}

\author[2]{Christoforos Rekatsinas}
\cormark[1]
\ead{crek@iit.demokritos.gr}

\affiliation[1]{organization={Department of Informatics, University of Piraeus},
    addressline={80 Karaoli \& Dimitriou Str.},
    city={Piraeus},
    postcode={185 34},
    state={Attica},
    country={Greece}}

\affiliation[2]{organization={Institute of Informatics and Telecommunications, National Centre for Scientific Research ``Demokritos''},
    addressline={Patr. Gregoriou E \& 27 Neapoleos Str},
    city={Agia Paraskevi},
    postcode={15341},
    state={Attica},
    country={Greece}}

\cortext[cor1]{Corresponding author}

\begin{abstract}
Molecular dynamics (MD) simulations are governed by explicit equations of motion, yet most neural approaches that accelerate or emulate MD rely on simulator-generated trajectories, forces, or energies for training. In this work we ask to what extent can physically meaningful molecular trajectories be recovered from the governing laws. We introduce the Differentiable Newtonian Molecular Solver (DINaMo), a physics-informed neural framework that represents molecular trajectories as differentiable functions of time and is trained exclusively through Newtonian dynamics, conservation laws, and analytic interaction potentials on a given equilibrated initial state. Unlike prior physics-informed MD formulations, DINaMo uses no simulator-generated trajectories, forces, velocities, or energies as supervisory targets. In Lennard--Jones argon systems, the learned trajectories reproduce short-time coordinate, energy, and structural observables, including in a larger and denser liquid-like setting where the radial distribution function is recovered. Although currently limited to short temporal horizons, the results indicate that physically meaningful molecular trajectories can emerge directly from physics-only supervision, supporting the feasibility of trajectory-unsupervised neural solvers for molecular dynamics.
\end{abstract}


\maketitle

\section{Introduction}\label{sec1}

Molecular dynamics simulations provide a powerful computational framework for studying molecular structure, dynamics, and thermodynamic behavior at atomistic resolution.\citep{Frenkel2002} They are used across materials science,\citep{Chandran2026} drug design,\citep{RajaramBaskaran2025} biology,\citep{Kaur2026} and soft matter,\citep{Ciccotti2019} offering insight into mechanisms that are difficult or inaccessible experimentally. Yet despite advances in hardware and high-performance computing, MD remains demanding for complex systems such as proteins, macromolecules, and functionalized materials, particularly when long timescales or large configurational spaces must be explored.

Conventional MD is a numerical initial-value problem. \citep{LeimkuhlerReich2004} Given the initial positions and velocities of the atoms and a law for the forces between them, the equations of motion determine how the system evolves. The solution is the trajectory: the positions and velocities of every atom as functions of time.\citep{Frenkel2002,AllenTildesley2017} A computer reconstructs this trajectory by advancing the state through a large number of small time steps. The steps are sequential, each depends on the state produced by the one before it, so the calculations cannot readily be spread along the time axis the way they can be spread across atoms.\citep{Gander2015} Reaching the timescales on which phenomena of interest unfold is therefore costly in a way that faster hardware alone does not resolve, which is what motivates methods that replace or augment explicit step-by-step integration with a learned alternative.

Artificial intelligence (AI) and machine learning (ML) have therefore been widely adopted to accelerate molecular discovery and simulation.\citep{Taniguchi2026} A common strategy is high-throughput screening, where a subset of candidate systems is simulated, a ML model is trained on the generated data, and predictions are extrapolated to the remaining design space.\citep{Sezgin2025} However, such approaches depend on sufficiently large and representative datasets. In large design spaces, randomly sampled training sets may miss rare high-performing candidates, compromising predictive reliability and limiting discovery performance.\citep{Krokidas2025} Beyond surrogate screening, alternative paradigms aim to accelerate the simulations themselves. These include generative and dynamical frameworks that bypass explicit integration by sampling equilibrium distributions or approximating dynamics, such as Boltzmann generators,\citep{Olsson2026} normalizing-flow and diffusion approaches for equilibrium sampling,\citep{Bonneau2026} neural perturbations of Langevin dynamics for transition-path discovery,\citep{Seong2024} stochastic trajectory-prediction methods trained on MD rollouts,\citep{Schreiner2023} and deterministic neural models trained to reproduce molecular trajectories.\citep{Park2026} Despite their differences, these methods largely depend on extensive simulation-generated training data, and their ability to generalize beyond the training domain is often limited.\citep{Bonneau2026}

When a model is trained by comparing its predictions with a trajectory already computed by a simulator,\citep{Schreiner2023,Park2026} we refer to this as \emph{trajectory supervision}. Since MD is governed by explicit equations of motion and interaction laws, this raises a central question for ML applied to MD: \emph{to what extent can physically meaningful molecular trajectories be learned directly from the underlying physics, without trajectory supervision?}

Physics-informed neural networks (PINNs) provide a framework for pursuing this route.\citep{Raissi2019} Rather than learning from examples of the solution, a PINN is trained by minimizing the residuals of the governing equations evaluated on its own predictions, so the differential equation itself, rather than a dataset, supplies the training signal. We call an objective \emph{data-supervised} if any of its terms compares the prediction against a simulator-generated quantity, such as a reference position, velocity, force, or energy, and \emph{physics-supervised} if it uses no such quantity, so that every term instead measures how far the prediction departs from a governing physical relation, whether the equation of motion or a conservation law implied by it, evaluated from the prediction and its given initial state. So defined the two are mutually exclusive, and adopting the PINN formalism does not by itself make an objective physics-supervised, since a data-matching term can sit alongside physics residuals in the same composite loss.

Recent work has begun to explore physics-informed neural formulations for MD \citep{Razakh2021,Pham2024}. PND incorporates equations of motion, boundary conditions, and conservation laws into the training objective, but remains supported by MD-generated data \citep{Razakh2021}. More recently, the neuromorphic spiking framework NP-SNN was presented as a data-free solver for Lennard--Jones systems \citep{Pham2024}. However, its objective contains an explicit position term that compares the predicted positions onto the simulator trajectory, so MD data still enters as a supervisory signal. These studies demonstrate the feasibility of physically constrained neural solvers for MD, but in both cases the solution remains anchored to simulator-generated data. A fully physics-driven formulation, in which trajectories emerge solely from the governing equations without trajectory, force, or energy labels, remains an open challenge.

In this work, we introduce DINaMo, a trajectory-unsupervised, physics-informed neural network for MD. Unlike existing neural MD approaches that rely on simulator-generated trajectories or rollout supervision,\citep{Razakh2021,Pham2024} DINaMo has a physics-supervised objective, learning molecular trajectories directly from the governing physical laws. It represents molecular motion as a single continuous differentiable function of time which obtains both velocities and accelerations by differentiating that function rather than from separate network outputs. Tying the kinematics to the predicted path, this way keeps them mutually consistent, so the network cannot satisfy a velocity or energy residual with motion inconsistent with the trajectory it actually predicts. The cost is that a small high-frequency error in the path is amplified when it is differentiated, which makes the velocity the hardest quantity to fit and carries that error into the energy. Representing the whole window at once also removes the integration timestep, so there is no stability limit to respect and no sequential accumulation of stepping error, and the difficulty instead reappears as an optimization problem in which the stiffness of the force field conditions the loss landscape.

Learning such a trajectory from physics alone calls for several mechanisms, each addressing part of the problem. A hard initial-condition ansatz splits the trajectory into a fixed analytic part that captures the short-time motion already determined by the initial state and the equations of motion, plus a learned correction, so the neural trajectory starts at exactly the given initial state and the network need only represent the nonlinear departure that develops later. An augmented-Lagrangian energy-rate constraint holds the total energy consistent by adapting its penalty as training proceeds rather than through a weight fixed by hand. Because the objective is solved as a single optimization over the window, two further mechanisms govern where optimization pressure is applied within the window. Causal weighting enforces the arrow of time, holding back pressure at later times while earlier residuals remain large, and late-stage residual mining then concentrates pressure on the few atoms and times whose errors would otherwise stay hidden in the average. No simulator-generated positions, velocities, forces, energies, or trajectories are used as training targets at any stage.

We evaluate DINaMo on Lennard--Jones argon in the microcanonical (NVE) ensemble, across systems of increasing difficulty at the 50- and 500-atom sizes also studied by Pham et al.~\citep{Pham2024}. We match the system, ensemble, and these sizes, but temporal windows, temperatures, and densities may differ, so any comparison is one of approach and scale rather than of identical conditions (experiment details are listed in Table~\ref{tab:run_parameters}). The experiments examine whether physically consistent trajectories can be obtained from physics supervision alone, whether coherent dynamics can be maintained over extended temporal windows, and whether the physics-supervised approach remains effective in larger and denser systems. DINaMo achieves physical consistency over temporal windows which are substantially longer than those we have found reported for comparable physics-informed approaches, in the longest case by roughly an order of magnitude. Although the present work remains focused on short-horizon dynamics in a specific ensemble, with a separate model trained for each initial condition, results indicate that accurate molecular trajectories, stable energy behavior, and meaningful structural observables can indeed emerge from physics-only supervision.

The remainder of the paper is organized as follows. Section~\ref{sec:methods} formulates the problem and describes DINaMo and its physics-only training, Section~\ref{sec:results} reports the experiments on Lennard--Jones argon, and Sections~\ref{sec:discussion} and~\ref{sec:conclusions} discuss the findings and conclude.

\section{Methodology}\label{sec:methods}

DINaMo represents the whole trajectory on the fixed window $[0,T]$ as a single differentiable function of time $\mathbf r_\theta(t)$, conditioned on an equilibrated initial state, and fixes its parameters $\theta$ by a single optimization over the window rather than by stepping the state forward. The optimizer must find a path whose second derivative matches a stiff many-body force field, while small errors in the path are amplified in the velocity, energy, and pair-structure observables. Its objective is built entirely from physics residuals, with no simulator quantity as a target, and is organized as a hierarchy: Newton's equation is the primary residual, momentum and energy conservation act as consistency checks, residuals are weighted causally in time, energy-rate drift is controlled by an augmented Lagrangian, and collocation is fixed on the reference MD time grid.

\begin{figure}[!t]
\centering
\includegraphics[width=\textwidth,height=1.0\textheight,keepaspectratio]{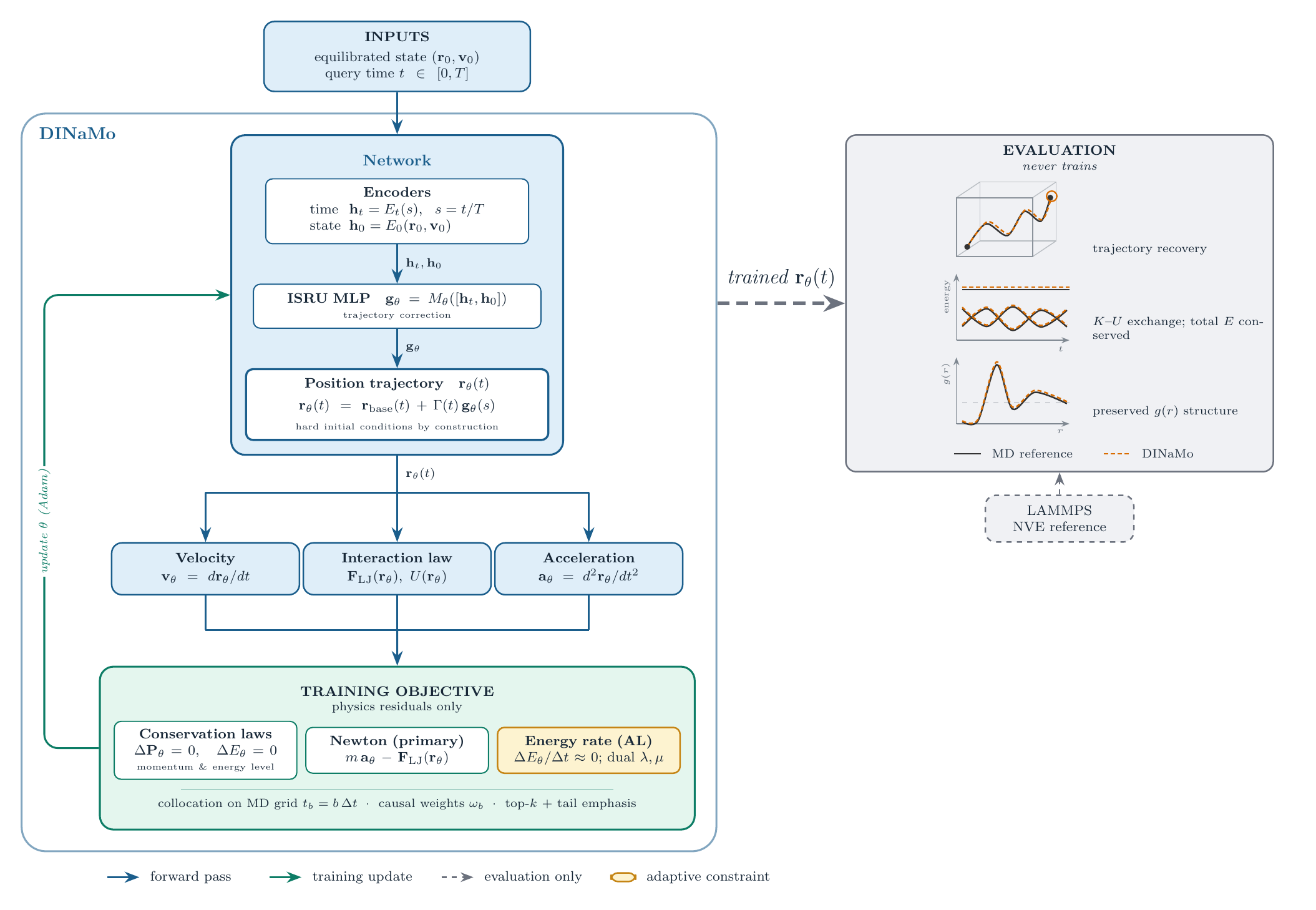}
\caption{Architecture and physics-only training of DINaMo. Time and initial-condition encoders condition a multilayer perceptron that predicts a position correction $\mathbf g_\theta$. A hard initial-value ansatz imposes the initial position and velocity by construction, and the $a_0$ variant shown also imposes the initial acceleration. Velocities and accelerations are derivatives of the resulting position trajectory, and Lennard--Jones forces are evaluated analytically. The objective combines a primary Newton residual with momentum and energy conservation and an augmented-Lagrangian energy-rate term. Collocation is fixed on the MD time grid, with ratio-based causal annealing and late-stage atom-time top-$k$ and tail emphasis. Every term is a physics residual built from the prediction and the analytic force. Reference MD trajectories are used only for evaluation and never during training.}
\label{fig:architecture_diagram}
\end{figure}

Figure~\ref{fig:architecture_diagram} summarizes the architecture and the physics-only training. The following subsections formalize the problem, then develop the trajectory representation and its hard initial-condition ansatz, the residual objective with its causal and late-stage weighting, the augmented-Lagrangian energy-rate constraint, and the training procedure in turn.

\subsection{Problem formulation}\label{sec:methods_problem}

We consider an $N$-atom system with positions $\mathbf r(t)=(\mathbf r_1(t),\ldots,\mathbf r_N(t))\in\mathbb R^{3N}$, velocities $\mathbf v(t)=\dot{\mathbf r}(t)$, and atomic mass $m$. The dynamics are governed by
\begin{equation}
m\ddot{\mathbf r}(t)=\mathbf F(\mathbf r(t)),
\qquad
\mathbf r(0)=\mathbf r_0,
\qquad
\dot{\mathbf r}(0)=\mathbf v_0.
\label{eq:newton_ivp}
\end{equation}
For a Lennard--Jones system,
\begin{equation}
U(\mathbf r)=\sum_{1\le i<j\le N}4\epsilon\left[\left(\frac{\sigma}{r_{ij}}\right)^{12}-\left(\frac{\sigma}{r_{ij}}\right)^6\right],
\qquad
\mathbf F(\mathbf r)=-\nabla_{\mathbf r}U(\mathbf r),
\label{eq:lj}
\end{equation}
with minimum-image pair distances $r_{ij}$. The pair interaction is a finite-range, shifted-force Lennard--Jones potential whose energy and force both vanish smoothly at the cutoff. The LAMMPS pair style and cutoff are given in Appendix~\ref{app:ic_grid}. The neural residual uses the same force convention as the reference data. This holds the network to exactly the force law that generated the comparison trajectory.

The total momentum and total energy of a predicted trajectory are
\begin{equation}
\mathbf P_\theta(t)=m\sum_{i=1}^{N}\mathbf v_{\theta,i}(t),
\qquad
E_\theta(t)=\frac{1}{2}m\sum_{i=1}^{N}\|\mathbf v_{\theta,i}(t)\|_2^2+U(\mathbf r_\theta(t)).
\label{eq:momentum_energy}
\end{equation}
We study the system in the microcanonical, or NVE, ensemble, in which the number of atoms $N$, the volume $V$, and the total energy $E$ are held fixed. Physically this is an isolated system that exchanges neither energy nor particles with its surroundings, so Newton's equations conserve the total energy, and in the periodic box used here they also conserve the total linear momentum. These conserved quantities are what the consistency residuals below monitor. The reference trajectories are equilibrated beforehand at a target temperature in the canonical (NVT) ensemble. That step and its coupling to DINaMo are described in Section~\ref{sec:methods_reference}.

\FloatBarrier

\subsection{Architecture and trajectory representation}\label{sec:methods_model}

DINaMo is a conditional continuous-time neural solver. Given a query time $t\in[0,T]$ and an initial state $(\mathbf r_0,\mathbf v_0)$, it returns a predicted position $\mathbf r_\theta(t)\in\mathbb R^{3N}$. It does not propagate the system autoregressively from one discrete MD step to the next. Instead, the whole short-time trajectory is represented by a single differentiable function. This makes the network a global trajectory representation for an initial-value problem rather than a learned time-stepper, and it allows velocities and accelerations to be obtained as exact derivatives.

The architecture has three functional parts (Fig.~\ref{fig:architecture_diagram}). A time encoder describes where the query lies within the horizon, an initial-condition encoder describes the molecular state being solved, and an inverse square root unit (ISRU)\citep{Ohn2019} multilayer perceptron maps the combined representation to a trajectory correction. The correction is then inserted into a hard initial-condition ansatz, so the network learns only the nonlinear departure from the local Newtonian expansion rather than the entire trajectory from scratch.

The representation is position-only:
\begin{equation}
\mathbf v_\theta(t)=\frac{d\mathbf r_\theta}{dt}(t),
\qquad
\mathbf a_\theta(t)=\frac{d^2\mathbf r_\theta}{dt^2}(t).
\label{eq:derived_kinematics}
\end{equation}
There are no independent velocity or acceleration output heads. Velocity and acceleration are not free quantities in Newtonian mechanics. Tying them to the position field by differentiation removes a way for the network to satisfy a velocity or energy residual with kinematics that are inconsistent with the path it actually predicts.

Let $s=t/T$ denote normalized time and let $\mathbf z_0=[\mathbf r_0,\mathbf v_0]\in\mathbb R^{6N}$. The network uses separate encoders for time and initial conditions,
\begin{equation}
\mathbf h_t=E_t(s),
\qquad
\mathbf h_0=E_0(\mathbf z_0),
\qquad
\mathbf y_\theta=M_\theta([\mathbf h_t,\mathbf h_0])\in\mathbb R^{3N},
\label{eq:architecture}
\end{equation}
where $E_t$ is a time encoder, $E_0$ is an initial-condition encoder, and $M_\theta$ is an ISRU multilayer perceptron. The vector $\mathbf y_\theta$ is interpreted as the correction $\mathbf g_\theta$ used in Eq.~\ref{eq:hard_ic_ansatz}.

\subsubsection{Hard initial-condition ansatz}\label{subsec:ic_enforcement}

The predicted trajectory is written as
\begin{equation}
\mathbf r_\theta(t)
=
\mathbf r_0+t\mathbf v_0+\frac{1}{2}t^2\mathbf a_0
+T^2s^3\mathbf g_\theta(s;\mathbf r_0,\mathbf v_0),
\qquad
\mathbf a_0=\frac{1}{m}\mathbf F_{\mathrm{LJ}}(\mathbf r_0).
\label{eq:hard_ic_ansatz}
\end{equation}
Here $s=t/T\in[0,1]$ is the normalized time and $\mathbf g_\theta$ is the network output of Eq.~\ref{eq:architecture}. Each part of this ansatz is chosen deliberately. The leading polynomial $\mathbf r_0+t\mathbf v_0+\tfrac{1}{2}t^2\mathbf a_0$ is not an arbitrary baseline but the exact second-order Taylor expansion of the true trajectory about $t=0$, because the governing dynamics fix the value, the first derivative, and the second derivative there: $\mathbf r(0)=\mathbf r_0$, $\dot{\mathbf r}(0)=\mathbf v_0$, and $\ddot{\mathbf r}(0)=\mathbf a_0=\mathbf F_{\mathrm{LJ}}(\mathbf r_0)/m$. The network therefore does not have to represent the analytically known short-time motion and only has to model the higher-order remainder that accumulates as the trajectory departs from this local expansion.

Expressing the correction in the normalized time $s$ rather than the physical time $t$ decouples the learned shape from the absolute length of the window: $\mathbf g_\theta(s)$ describes the departure as a fraction of the horizon, which keeps the optimization comparably conditioned when $T$ is changed between runs. The cubic gate $s^3$ is the lowest integer power for which appending the correction leaves all three analytically fixed initial quantities untouched. A factor $s^1$ would corrupt the initial velocity and $s^2$ the initial acceleration, whereas $s^3$, together with its first two time derivatives, vanishes at $t=0$, so the initial constraints hold by construction rather than as penalties:
\begin{equation}
\mathbf r_\theta(0)=\mathbf r_0,
\qquad
\dot{\mathbf r}_\theta(0)=\mathbf v_0,
\qquad
\ddot{\mathbf r}_\theta(0)=\mathbf a_0.
\label{eq:hard_constraints}
\end{equation}
The prefactor $T^2$ restores the physical scale removed by the normalization. Since $T^2s^3=t^2s$, the correction is the familiar $t^2$ ballistic weight ramped smoothly from zero by the factor $s$, and it carries the dimensions of length precisely when $\mathbf g_\theta$ is of order the system's acceleration scale, matching the $\tfrac{1}{2}t^2\mathbf a_0$ term it refines. Keeping $\mathbf g_\theta$ at order unity in these units, rather than letting it span many orders of magnitude across the window, improves the conditioning of the optimization. The practical impact is twofold: three soft initial-condition penalties that would otherwise compete with the Newton residual are removed entirely, and optimizer capacity is concentrated on the genuinely hard part of the problem, the late-time nonlinear motion.

The same baseline-plus-gated-correction construction admits two further variants. A simpler first-order form omits the initial-acceleration term and gates the correction by $s^2$,
\begin{equation}
\mathbf r_\theta(t)=\mathbf r_0+t\mathbf v_0+T^2s^2\,\mathbf g_\theta(s;\mathbf r_0,\mathbf v_0),
\label{eq:ansatz_simple}
\end{equation}
which still enforces $\mathbf r_\theta(0)=\mathbf r_0$ and $\dot{\mathbf r}_\theta(0)=\mathbf v_0$ exactly but leaves the initial acceleration to the network. Here $s^2$ is the lowest power that preserves the two fixed conditions, by the same argument as above. A bounded warp-ballistic form instead keeps $\mathbf r_0$ and $\mathbf v_0$ exact while replacing the unbounded straight-line displacement $t\mathbf v_0$ with a saturating one,
\begin{equation}
\mathbf r_\theta(t)=\mathbf r_0+\tau_{\mathrm w}\tanh\!\left(\frac{t}{\tau_{\mathrm w}}\right)\mathbf v_0+t^2\,\mathbf g_\theta(t;\mathbf r_0,\mathbf v_0).
\label{eq:ansatz_bounded}
\end{equation}
Because $\tau_{\mathrm w}\tanh(t/\tau_{\mathrm w})\to t$ as $t\to0$ and saturates at $\tau_{\mathrm w}$ for $t\gg\tau_{\mathrm w}$, the leading displacement is ballistic at early times yet bounded by $\tau_{\mathrm w}\lVert\mathbf v_0\rVert$ thereafter. This keeps the analytic baseline from carrying atoms arbitrarily far in a straight line over longer or denser windows, leaving the network only the smaller bending of the trajectory to represent rather than a large and growing ballistic displacement. All three share the property that the enforced initial conditions hold by construction rather than through penalty terms, differing only in how many of them are fixed. In the reported experiments, the $a_0$ form is used for the base run, while the bounded warp-ballistic form with $\tau_{\mathrm w}=0.12$ is used for the extended and dense runs. In the bounded form, the initial acceleration is enforced softly by the Newton residual rather than by construction, while the saturating baseline keeps the leading ballistic displacement bounded over longer or denser settings. The simpler first-order form of Eq.~\ref{eq:ansatz_simple} is a third option within the same family that is not used in the runs reported here.

\subsubsection{ISRU nonlinearity}\label{subsec:isru}

Hidden layers use a smooth inverse-square-root activation
\begin{equation}
\phi(x)=\frac{x}{\sqrt{1+\alpha x^2}},
\qquad
\phi'(x)=(1+\alpha x^2)^{-3/2},
\qquad
\phi''(x)=-3\alpha x(1+\alpha x^2)^{-5/2}.
\label{eq:isru}
\end{equation}
This activation is chosen because the loss requires first and second derivatives of the network output with respect to time. Since $\mathbf a_\theta$ is a second derivative of the network, the activation must itself have well-behaved first and second derivatives. The ISRU form is smooth and bounded with closed-form $\phi'$ and $\phi''$, avoiding the exploding or vanishing higher-order derivatives that piecewise-linear or hard-saturating activations introduce when differentiated twice.

\subsection{Residual loss formulation}\label{sec:methods_loss}

Training minimizes physical residuals at fixed collocation times $\mathcal T_B=\{t_b\}_{b=1}^{B}\subset[0,T]$, the discrete times within the window at which the governing-equation residuals are evaluated during training. These are indexed by the same $b$ that labels the reference MD frames, since the collocation grid is placed on the MD time grid (Section~\ref{sec:methods_collocation}). At each $t_b$, the model evaluates $\mathbf r_\theta(t_b)$, obtains $\mathbf v_\theta(t_b)$ and $\mathbf a_\theta(t_b)$ by automatic differentiation, evaluates Lennard--Jones forces and energies from the predicted positions, and forms per-time residuals. Every residual below is a function of the prediction and the analytic force only. No simulator value is used as a target.

The Newton residual is
\begin{equation}
N_b=\frac{1}{N}\sum_{i=1}^{N}\left\|m\mathbf a_{\theta,i}(t_b)-\mathbf F_i(\mathbf r_\theta(t_b))\right\|_2^2.
\label{eq:loss_newton}
\end{equation}
Momentum and per-atom energy residuals are
\begin{equation}
P_b=\left\|\mathbf P_\theta(t_b)-\mathbf P_\theta(0)\right\|_2^2,
\qquad
E_b=\left(\frac{E_\theta(t_b)-E_\theta(0)}{Ns_E}\right)^2,
\label{eq:loss_momentum_energy}
\end{equation}
where $s_E$ is a fixed per-atom energy scale. Because the centre-of-mass velocity is removed when the initial state is prepared (Section~\ref{sec:methods_reference}), $\mathbf P_\theta(0)\approx\mathbf 0$, so the momentum residual chiefly suppresses any spurious net momentum that would otherwise develop over the window. Removing the centre-of-mass velocity also fixes a common inertial frame for the neural and reference trajectories, so trajectory comparisons are not contaminated by a uniform drift of the whole box. The energy-rate residual is defined from the per-atom time derivative of the total energy,
\begin{equation}
W_b=\left(\frac{1}{s_E/T}\frac{1}{N}\frac{dE_\theta}{dt}(t_b)\right)^2.
\label{eq:loss_power}
\end{equation}
Differentiating the total energy of Eq.~\ref{eq:momentum_energy} and using $\mathbf F=-\nabla U$ gives $dE_\theta/dt=\sum_{i}\mathbf v_{\theta,i}\cdot(m\mathbf a_{\theta,i}-\mathbf F_i)$, the inner product of the atomic velocities with the corresponding Newton residuals. The energy rate therefore vanishes whenever Newton's equation is satisfied, so $W_b$ penalizes the energy-changing part of the dynamical error. It complements the level residual $E_b$ of Eq.~\ref{eq:loss_momentum_energy}: $W_b$ keeps the total energy from changing at each collocation time, whereas $E_b$ penalizes its accumulated departure from the value at $t=0$. Both residuals are normalized per atom by the fixed energy scale $s_E$, so the same weights apply across system sizes. Rather than treating $W_b$ as another fixed-weight soft penalty, the method handles energy-rate consistency through the augmented Lagrangian described below.

The fixed residual part of the objective is
\begin{equation}
R_b=w_NN_b+w_PP_b+w_EE_b.
\label{eq:fixed_residual}
\end{equation}
The weights are set before training to preserve a physical hierarchy: Newton's equation is the primary residual, while momentum and energy conservation act as consistency regularizers. Fixing this hierarchy in advance, rather than letting an adaptive scheme equalize the terms, is a deliberate choice. It keeps the equation of motion dominant and prevents the auxiliary residuals from quietly taking over the gradient when they happen to be large. Notably, all three terms compare the prediction to physical relations rather than to simulator data, so this is the point at which DINaMo departs from objectives that include a data-matching loss.

\subsubsection{Causal annealing}\label{sec:methods_causal}

Causal weighting plays a central role in the physics-only objective: without a temporal ordering, the optimizer can, in principle, lower the mean residual by fitting late times against an early trajectory that is not yet correct, yielding a solution that violates the arrow of time. To respect the temporal structure of the initial-value problem, residuals are therefore causally weighted. Let $D_b$ be a nonnegative driver residual formed from normalized Newton residuals, with a small admixture of the normalized energy and energy-rate residuals in the runs reported here. The causal weight is
\begin{equation}
\omega_b=\frac{B\exp\left(-\epsilon_c\sum_{j<b}D_j\right)}{\sum_{\ell=1}^{B}\exp\left(-\epsilon_c\sum_{j<\ell}D_j\right)}.
\label{eq:causal_weight}
\end{equation}
Large early residuals therefore reduce the weight of later times until the early segment improves. The intuition mirrors the arrow of time in an initial-value problem: a later state cannot be correct before the states that produce it are, so optimization pressure at $t_b$ is suppressed while earlier residuals remain large.

The coefficient $\epsilon_c$ is annealed using the ratio between late-window and early-window driver residuals. If the tail-to-head ratio is too large, $\epsilon_c$ is decreased so that later times receive more direct optimization pressure. Otherwise, $\epsilon_c$ is increased up to a fixed maximum to maintain causal ordering. This ratio-based annealing differs from a static causal mask. The schedule responds to how far the solution has actually progressed through the window rather than following a fixed curve. Because the adjustment is driven by the measured tail-to-head residual ratio rather than a preset schedule, the annealing rebalances optimization pressure dynamically as training proceeds, and the same settings adapt automatically to runs that converge at very different rates. In our runs, a single configuration handled the heterogeneous initial conditions of Section~\ref{sec:robustness} without per-run adjustment of these settings.

\subsubsection[Top-k and tail residual emphasis]{Top-$k$ and tail residual emphasis}\label{sec:methods_topk_tail}

After the main trajectory has converged, a small number of localized residuals can be hidden by the global mean. The final stage therefore applies atom-time top-$k$ mining to the Newton residual, selecting the largest residuals across atoms and collocation times and adding their mean to the objective. In addition, the final portion of the time window receives an explicit tail residual term. Once the average residual is small, a few atoms at a few late times can still carry most of the remaining error while being invisible to the global average. These two mechanisms put optimization pressure exactly where the dominant failure mode lives. As Section~\ref{sec:t020_results} shows, that failure mode is the late-time derivative error that is small on average but visible in the maximum velocity error.

\subsection{Augmented-Lagrangian energy-rate constraint}\label{sec:methods_al}

The energy-rate constraint is enforced through the average normalized residual
\begin{equation}
c(\theta)=\frac{1}{B}\sum_{b=1}^{B}W_b.
\label{eq:al_constraint}
\end{equation}
During the initial warmup, this acts as a small soft penalty. After warmup, the augmented-Lagrangian term~\citep{Hestenes1969,Powell1969} is
\begin{equation}
\mathcal L_{\mathrm{AL}}(\theta)=\lambda c(\theta)+\frac{\mu}{2}c(\theta)^2.
\label{eq:al_loss}
\end{equation}
The dual variable and penalty parameter are then updated outside the gradient graph,
\begin{equation}
\lambda\leftarrow\max\{0,\lambda+\mu c(\theta)\},
\qquad
\mu\leftarrow\min\{\mu_{\max},\gamma_\mu\mu\},
\label{eq:al_update}
\end{equation}
at prescribed intervals. The updates in Eq.~\ref{eq:al_update} are a dual-ascent step on the constraint. The multiplier $\lambda$ accumulates the running constraint violation across training, so a violation the optimizer repeatedly fails to remove raises $\lambda$ and, through the linear term $\lambda c(\theta)$, applies steadily increasing pressure, while the penalty coefficient $\mu$ is grown geometrically up to the cap $\mu_{\max}$ to sharpen enforcement. This is the advantage over a single fixed-weight penalty: a pure quadratic penalty has to be made very large to enforce the constraint, at which point it can overwhelm the Newton residual, whereas the multiplier enforces energy-rate consistency at finite $\mu$ and adapts the effective penalty as training proceeds rather than through a weight chosen by hand. The impact is a controlled total energy across the window, as seen in the energy panels of Fig.~\ref{fig:energy_all}.

\subsection{Training procedure}\label{sec:methods_training}

The full procedure is summarized in Algorithm~1 and, schematically, in the training loop of Fig.~\ref{fig:architecture_diagram}.

\begin{center}
\setlength{\fboxsep}{10pt}
\fbox{%
\begin{minipage}{0.92\linewidth}
\textbf{Algorithm 1: Physics-only training of DINaMo}

\textbf{Input:} Initial state $(\mathbf r_0,\mathbf v_0)$, mass $m$, force law $\mathbf F_{\mathrm{LJ}}$, training horizon $T$, MD time step $\Delta t$, epoch budget $E$, initial network parameters $\theta$.

\textbf{Output:} Trained parameters $\theta$ defining $\mathbf r_\theta(t)$.

\begin{enumerate}
    \item Construct the fixed collocation grid $t_b=b\Delta t$ on $[0,T]$.
    \item \textbf{for} epoch $=1,\dots,E$ \textbf{do}
    \begin{enumerate}
        \item Evaluate $\mathbf r_\theta(t_b)$ at all collocation times.
        \item Differentiate to obtain $\mathbf v_\theta(t_b)$ and $\mathbf a_\theta(t_b)$.
        \item Compute Lennard--Jones forces, momentum, total energy, and energy rate.
        \item Form fixed-weight Newton, momentum, and energy residuals.
        \item Apply causal weights and update the causal coefficient by tail-to-head annealing.
        \item Apply late-stage top-$k$ residual mining and tail emphasis when active.
        \item Add the augmented-Lagrangian energy-rate term.
        \item Update $\theta$ with Adam. At AL update intervals, update $\lambda$ and $\mu$.
    \end{enumerate}
    \item \textbf{end for}
\end{enumerate}
\end{minipage}
}
\end{center}

Each run is trained for a fixed number of epochs, rather than with an automatic stopping rule, and a single budget is not used across all systems. Because initial conditions differ in difficulty, and because the residual magnitudes stabilize sooner for some systems than others, we set the epoch budget per system, giving harder or longer-horizon settings more epochs. In each case the budget was chosen so that the Newton- and energy-residual magnitudes had levelled off by the end of training (Section~\ref{sec:robustness}). The epoch budgets used for the runs reported here are listed in Table~\ref{tab:run_parameters}.

\subsection{Fixed grid collocation}\label{sec:methods_collocation}

Collocation times are placed on the MD reference grid,
\begin{equation}
t_b=b\Delta t,
\qquad
b=0,1,\ldots,\left\lfloor\frac{T}{\Delta t}\right\rfloor.
\label{eq:fixed_grid}
\end{equation}
For the reported $T=0.12$ runs with $\Delta t=5\times10^{-4}$, this gives $B=241$ collocation points. For the $T=0.20$ run it gives $B=401$ points. Keeping the grid fixed makes each residual location persistent across epochs. This persistence allows causal weighting and late-window error repair to act on a stable set of points rather than chasing a resampled cloud. The difficulty that motivated this choice is described in Appendix~\ref{app:formulation_changes}.

\subsection{MD simulations}\label{sec:methods_reference}

The reference molecular-dynamics simulations are performed with LAMMPS~\citep{Thompson2022}, a widely used and extensively validated open-source classical MD engine, in reduced Lennard--Jones units. Lengths are measured in the pair diameter $\sigma$, energies in the well depth $\epsilon$, mass in the atomic mass $m$, time in $\tau_{\mathrm{LJ}}=\sigma\sqrt{m/\epsilon}$, temperature in $\epsilon/k_B$, and number density in $\sigma^{-3}$. The runs set $\epsilon=\sigma=m=1$. Each system is a cubic, fully periodic box of side $L=(N/\rho)^{1/3}$ holding $N$ argon-like atoms at number density $\rho$, interacting through a finite-range, shifted-force Lennard--Jones potential whose energy and force vanish smoothly at the cutoff (the exact LAMMPS pair style and cutoff are those of Section~\ref{sec:methods_problem} and are restated in Appendix~\ref{app:ic_grid}). After an overlap-guarded build and a short bad-contact relaxation, the configuration is energy-minimized, assigned Maxwell--Boltzmann velocities at the target reduced temperature $\Theta$\footnote{The two reduced temperatures are chosen on different bases. For the $N=50$ systems, $\Theta=2.5049$ corresponds to a room-temperature argon target of about $300$~K, using the standard argon well depth $\epsilon/k_B\approx119.8$~K, converted to reduced units. For the $N=500$ system, $\Theta=0.70$ corresponds to about $84$~K, a reduced dense-liquid state point specified directly. The $50$-atom runs are therefore dilute and supercritical (critical temperature $T_c^\ast\approx1.3$) and show the gas-like $g(r)$ of Fig.~\ref{fig:rdf_all}, whereas the $500$-atom run is a dense liquid, so cross-run temperature comparisons are not meaningful.} with the net linear and angular momentum removed, and equilibrated in the canonical (NVT) ensemble. Production trajectories are then generated in the microcanonical (NVE) ensemble with the velocity--Verlet integrator at timestep $\Delta t=5\times10^{-4}\,\tau_{\mathrm{LJ}}$, storing positions, velocities, and energies on the MD time grid. Conversion of the reduced quantities to SI units is given in Appendix~\ref{app:si_conversion}. Simulation and initial-state details are given in Appendix~\ref{app:ic_grid} and in the accompanying code repository.

DINaMo is coupled to these simulations only through the equilibrated initial state. It receives the post-equilibration positions and velocities, the box geometry, the atomic mass, the analytic interaction law, and the training collocation times. It receives no subsequent NVE state or observable as a training target. The NVE positions, velocities, forces, and energies are withheld during training and used only afterwards to compute the trajectory, energy, and radial-distribution comparisons of Section~\ref{sec:results}. We therefore describe the method as trajectory-unsupervised rather than data-free. It is not information-free. The initial state, mass, box, interaction law, and collocation times that define and discretize the molecular initial-value problem are still required, but no quantity from the subsequent reference rollout is used as a training target.

\section{Results}\label{sec:results}

DINaMo learns a continuous-time molecular trajectory for a Lennard--Jones system in the microcanonical (NVE) ensemble directly from the initial state, Newton's equation, conservation laws, and an analytic Lennard--Jones force law. It is not trained to reproduce a simulator rollout. Given a query time $t\in[0,T]$ and an initial state $(\mathbf r_0,\mathbf v_0)$, the model returns predicted atomic positions $\mathbf r_\theta(t)$ (where $\theta$ denotes network parameters), and velocities and accelerations are obtained by differentiating this position trajectory with respect to time. The complete architecture and the physics-only training loop are summarized in Fig.~\ref{fig:architecture_diagram}.

The experiments are arranged by increasing difficulty. The base experiment asks whether the architecture can recover an accurate short-time trajectory for a benchmark $50$-atom Lennard--Jones system. The second experiment extends the time window directly and asks whether physical coherence survives once trajectory errors have grown. The third experiment increases system size and number density, testing whether the same physics-only signal remains useful when local packing and force variation become more demanding.

\begin{figure}[p]
\centering
\includegraphics[width=\linewidth,height=0.20\textheight,keepaspectratio]{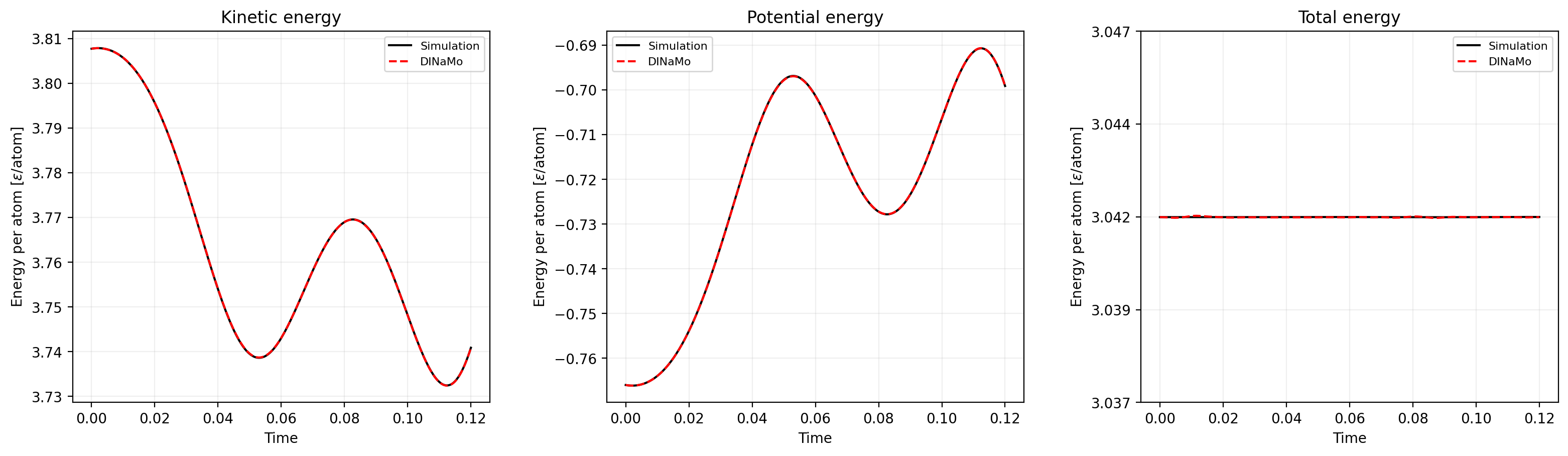}\\[2pt]
{\footnotesize (a)}\\[5pt]
\includegraphics[width=\linewidth,height=0.20\textheight,keepaspectratio]{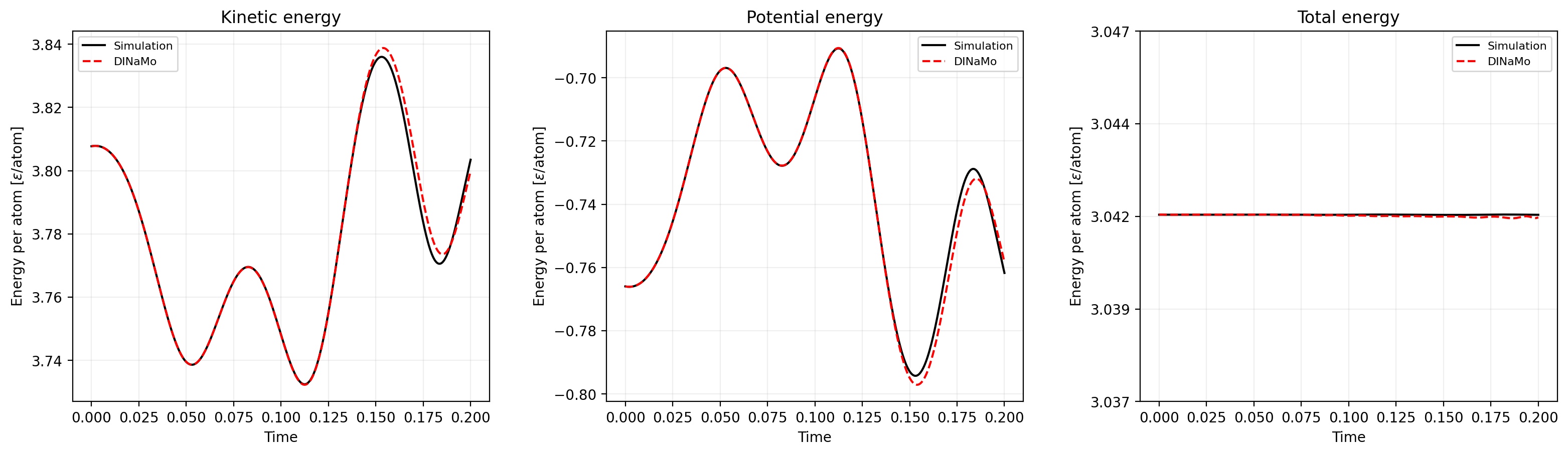}\\[2pt]
{\footnotesize (b) }\\[5pt]
\includegraphics[width=\linewidth,height=0.20\textheight,keepaspectratio]{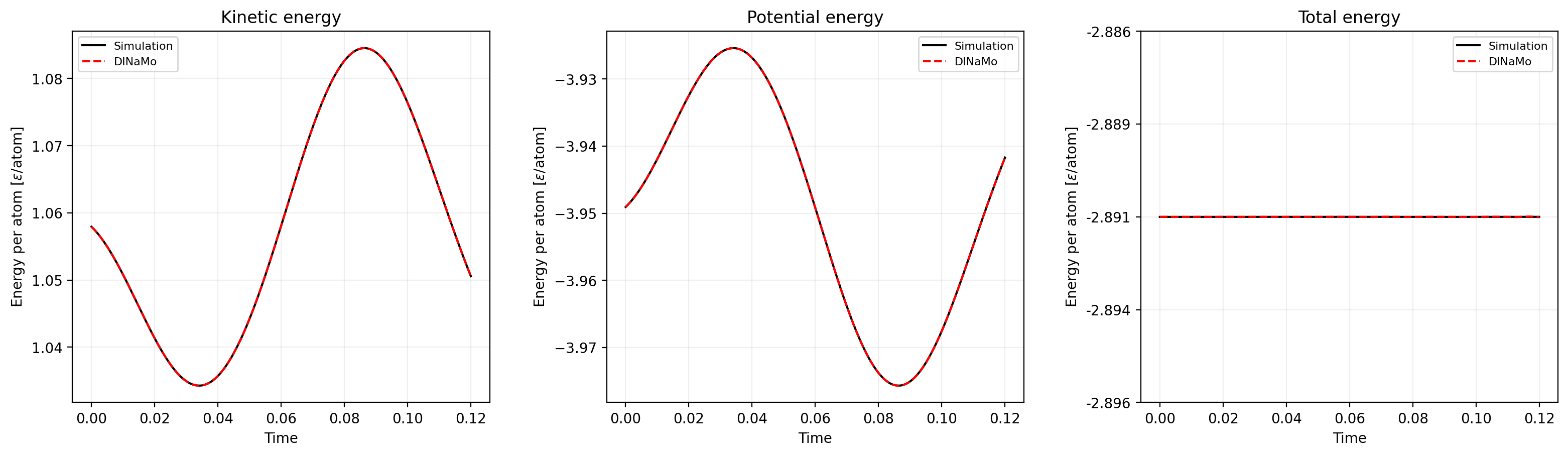}\\[2pt]
{\footnotesize (c)}
\caption{Kinetic, potential, and total energy versus time for the neural trajectory (dashed) and the LAMMPS reference (solid). Panel (a) shows the base trained horizon, $N=50$, $\rho=0.15$, $T=0.12$. Panel (b) shows the direct horizon extension, $N=50$, $\rho=0.15$, $T=0.20$. Panel (c) shows dense scaling, $N=500$, $\rho=0.50$, $T=0.12$. Reference energies are never supplied to the model, so the agreement follows from the differential physics and the augmented-Lagrangian energy-rate constraint. Total energy remains bounded near the reference in every case. Energies are shown per atom in units of $\epsilon/\mathrm{atom}$. For consistent visualization across the three DINaMo cases, each total-energy panel uses a common full-width window of $0.010\,\epsilon/\mathrm{atom}$. The smaller residual deviations are quantified independently in Table~\ref{tab:energy_errors}. Reference energies are taken from LAMMPS's internal energy output. Additional position and velocity error heatmaps are collected in Appendix~\ref{app:error_heatmaps}.}
\label{fig:energy_all}
\end{figure}

\begin{figure}[p]
\centering
\includegraphics[width=0.62\linewidth,height=0.20\textheight,keepaspectratio]{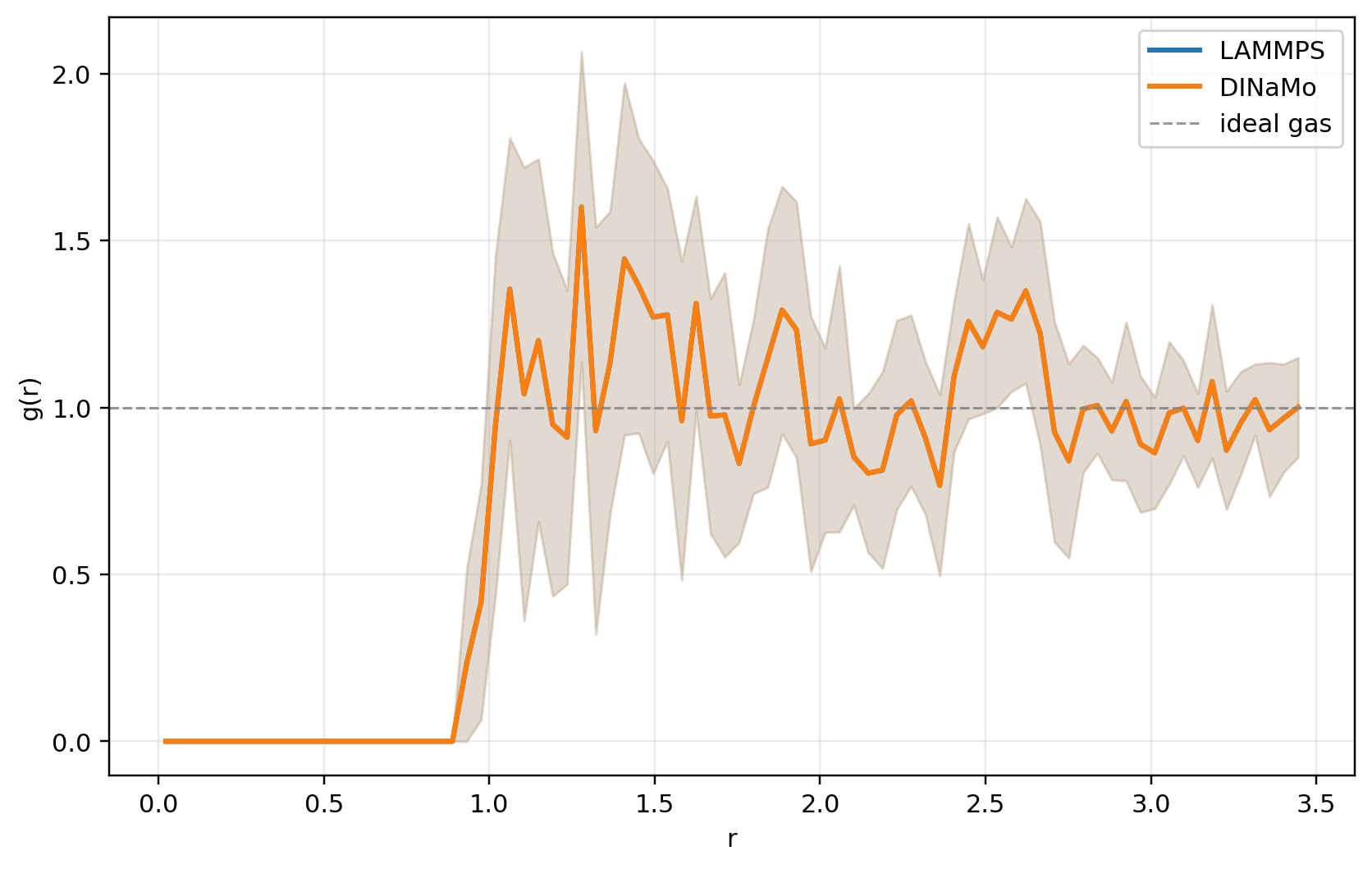}\\[2pt]
{\footnotesize (a) }\\[5pt]
\includegraphics[width=0.62\linewidth,height=0.20\textheight,keepaspectratio]{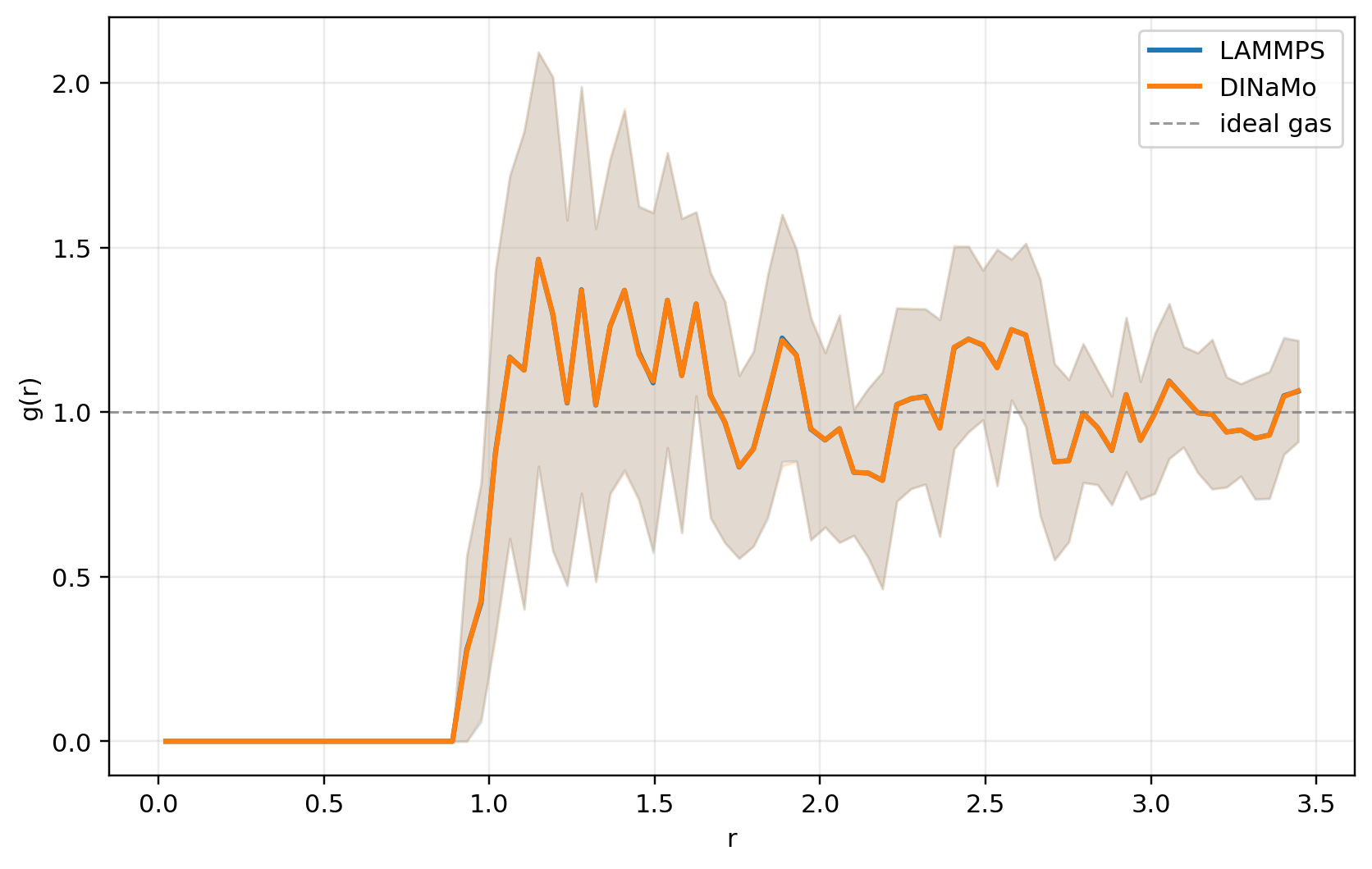}\\[2pt]
{\footnotesize (b) }\\[5pt]
\includegraphics[width=0.62\linewidth,height=0.20\textheight,keepaspectratio]{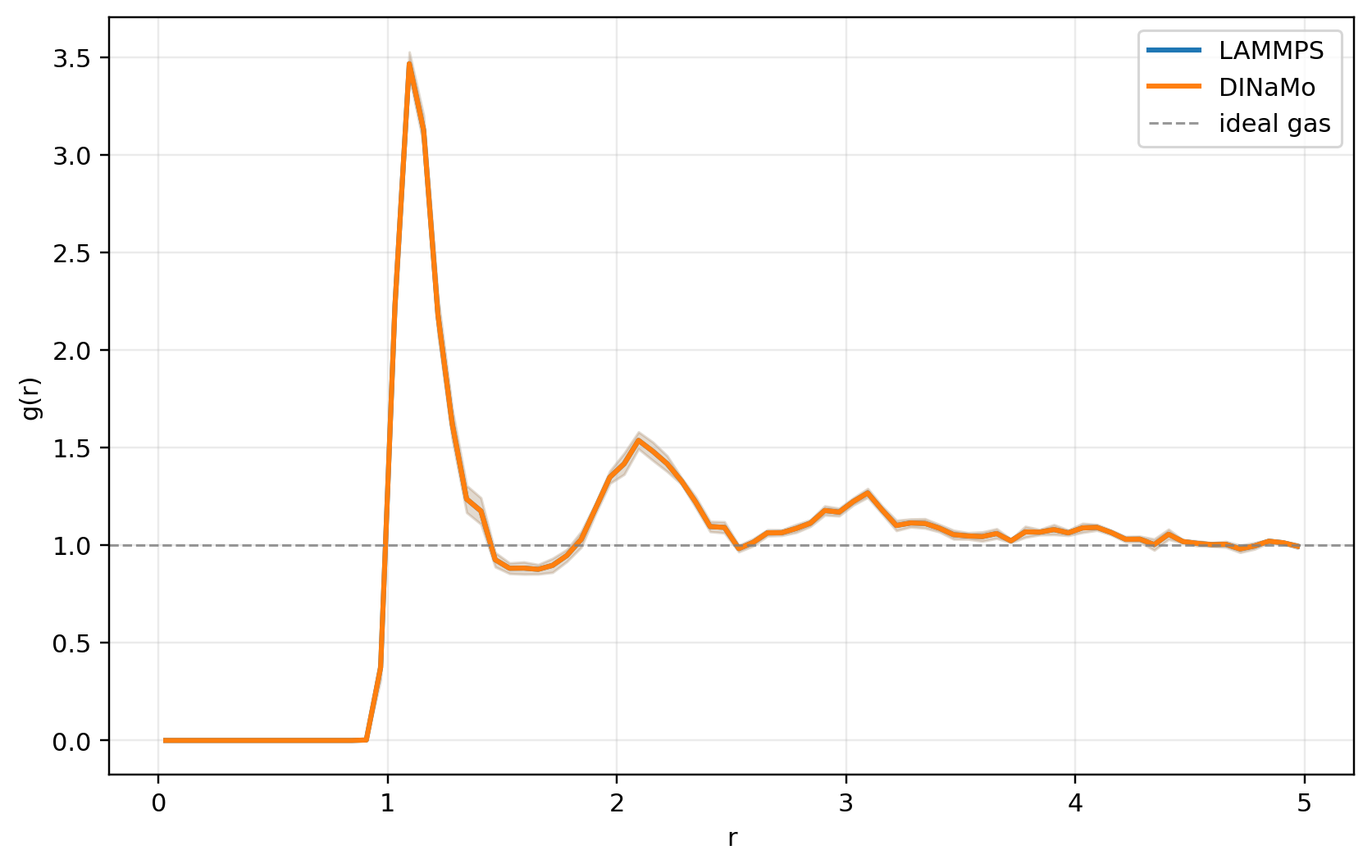}\\[2pt]
{\footnotesize (c) }
\caption{Radial distribution function $g(r)$ for the neural trajectory and the LAMMPS reference. Panel (a) shows the base low-density case, $N=50$, $\rho=0.15$, $T=0.12$. Panel (b) shows the direct horizon extension, $N=50$, $\rho=0.15$, $T=0.20$. Panel (c) shows the dense liquid-like case, $N=500$, $\rho=0.50$, $T=0.12$. The $50$-atom systems are nearly structureless and gas-like, whereas the dense $500$-atom system shows a pronounced first-neighbour peak ($g(r)\approx 3.5$) and liquid-like ordering. The neural $g(r)$ tracks the reference in all cases, reproducing structural information that is never supplied as a label. Curves show the time-averaged $g(r)$. The shaded band is $\pm 1$ standard deviation of the per-frame $g(r)$ across the averaged frames, and is wide for the $50$-atom systems simply because few pairs contribute to each bin in any single frame. The dashed horizontal line marks the ideal-gas limit $g(r)=1$. Additional position and velocity error heatmaps are collected in Appendix~\ref{app:error_heatmaps}.}
\label{fig:rdf_all}
\end{figure}

We evaluate each learned trajectory at three levels. Pointwise position and velocity errors test trajectory accuracy. Energy profiles test whether the predicted motion remains conservative. The radial distribution function $g(r)$ tests whether the local structure is preserved. None of these quantities is supplied as a training label. Figures~\ref{fig:energy_all} and~\ref{fig:rdf_all} show the energy and structural comparisons, while Tables~\ref{tab:heatmap_errors} and~\ref{tab:energy_errors} summarize the principal trajectory and energy errors. Sensitivity to the initial condition is examined separately in Section~\ref{sec:robustness}. All reported errors are evaluated over the complete reference-matched time grid in reduced Lennard--Jones units.

\subsection{Evaluation protocol}\label{sec:evaluation_protocol}

Reference trajectories are generated using the LAMMPS molecular-dynamics package~\citep{Thompson2022}: each system is energy-minimized and then equilibrated in the canonical (NVT) ensemble, after which the microcanonical (NVE) reference trajectory starts from that post-equilibration state (Section~\ref{sec:methods_reference} of Methods). During training, the neural solver receives only this initial state, which implicitly supplies the atom count $N$, the box geometry, atomic mass, and the analytic Lennard--Jones interaction. The subsequent NVE positions, velocities, forces, and energies are withheld and used only after training, for evaluation. Throughout, this LAMMPS rollout is treated as the numerical ground truth against which the neural trajectory is judged. This separation is what makes the comparisons below a test of physics-only learning rather than of supervised regression.

We assess three complementary observable layers. The first is pointwise trajectory accuracy over the full output trajectory. At each reference frame $t_b=b\Delta t$ (the $b$-th saved MD frame, with $\Delta t$ the simulation timestep), we compute per-atom position and velocity errors using the Euclidean ($\ell_2$) norm of the difference between the predicted and reference vectors for that atom:

\begin{equation}
e^{(r)}_i(t_b)=\left\|\mathbf r_{\theta,i}(t_b)-\mathbf r_{\mathrm{MD},i}(t_b)\right\|_2,
\qquad
e^{(v)}_i(t_b)=\left\|\mathbf v_{\theta,i}(t_b)-\mathbf v_{\mathrm{MD},i}(t_b)\right\|_2.
\label{eq:evaluation_errors}
\end{equation}

Using those metrics for the first layer, we report three trajectory-level summaries for each field. The first is the 95th-percentile trajectory error. At each saved frame we compute the 95th percentile over all atoms and then report the largest value over the time grid. The 95th percentile is a robust near-worst-case measure that is not skewed by a single outlier atom. The second is the trajectory maximum error, defined as the largest atom-norm error over all atoms and saved frames. We additionally report the overall median of the per-atom error norms over all atoms and saved frames. These quantities provide a good picture of accuracy over the entirety of a predicted trajectory.

The second layer is physical fidelity, measured through the time evolution of the kinetic, potential, and total energy over the window, compared against the reference. The third is structural fidelity, measured through the radial distribution function $g(r)$. This function gives the average pair density at separation $r$, normalized by the ideal-gas value, so that $g(r)=1$ indicates no positional correlation while peaks mark preferred neighbour separations. As the standard fingerprint that distinguishes gas-like from liquid-like local order, $g(r)$ probes whether the learned trajectory preserves molecular structure rather than merely matching coordinates in aggregate. Among these layers, the accuracy of velocity over a trajectory is the most stringent. Velocities are derivatives of the predicted positions and therefore amplify any high-frequency defect in the learned curve.

Reference trajectories are stored with 12 significant digits, so file readout precision does not limit the reported trajectory statistics. Reference energies in Table~\ref{tab:energy_errors} and Fig.~\ref{fig:energy_all} are taken directly from LAMMPS's internal double-precision accounting. For the principal runs, energies recomputed from the stored positions and velocities agree with that internal accounting to better than $2\times10^{-10}\,\epsilon$ per atom.

The exact configuration of each run is listed in Table~\ref{tab:run_parameters}.

\subsection[Accurate trained horizon at T=0.12]{Accurate trained horizon at $T=0.12$}\label{sec:t012_results}

The base run establishes the high-accuracy regime. Over the complete interval $t\in[0,0.12]$, the near-worst-case position error remains below $5\times10^{-5}\,\sigma$, and the maximum velocity error remains below $2\times10^{-3}\,\sigma/\tau_{\mathrm{LJ}}$ (Table~\ref{tab:heatmap_errors}). Every energy component stays within $10^{-4}\,\epsilon$ per atom of the reference. The model total-energy drift is $3.5\times10^{-5}\,\epsilon$ per atom, compared with $4.2\times10^{-6}\,\epsilon$ per atom for the velocity--Verlet reference (Table~\ref{tab:energy_errors}).

The energy profile provides a check beyond coordinate fitting alone. As shown in Fig.~\ref{fig:energy_all}a, kinetic and potential energies follow the LAMMPS reference throughout the window, and the total energy remains close to the reference trajectory. The kinetic and potential components continue to exchange energy, so the constraint has not flattened the profile to an artificial constant. Because the reference energy is never supplied to the model, this agreement is a consequence of satisfying the differential physics and the energy-rate constraint, not of supervised regression. The radial distribution function is likewise preserved (Fig.~\ref{fig:rdf_all}a). Because $g(r)$ aggregates information over many atom pairs and time points, agreement in $g(r)$ indicates that the neural trajectory retains the same short-range structural statistics as the reference dynamics rather than averaging out coordinate errors. In short, at the base setting the learned trajectory matches the reference in coordinates, energy, and structure simultaneously.

\subsection[Direct horizon extension to T=0.20]{Direct horizon extension to $T=0.20$}\label{sec:t020_results}

For the second experiment, we retained the base network width, depth, optimizer, and collocation strategy, but used the bounded warp-ballistic ansatz of Eq.~\ref{eq:ansatz_bounded} with $\tau_{\mathrm w}=0.12$ and trained directly on the longer window $T=0.20$. Increasing the network width or depth did not improve accuracy on this window, so the $256$-unit, three-layer network was retained (Table~\ref{tab:run_parameters}). This run is less accurate than the $T=0.12$ base case, but it remains consistent in the sense most relevant for a physics-only solver: energy components retain the correct qualitative evolution and the total energy stays bounded near the reference (Fig.~\ref{fig:energy_all}b), and the RDF continues to match the reference structural statistics (Fig.~\ref{fig:rdf_all}b).

The longer horizon changes the error mode more than it changes the conservation behavior. The trajectory maximum position error rises to $3.02\times10^{-3}\,\sigma$, and the maximum velocity error reaches $1.57\times10^{-1}\,\sigma/\tau_{\mathrm{LJ}}$ (Table~\ref{tab:heatmap_errors}). For position and velocity separately, only two of the fifty atoms have peak errors above half the corresponding trajectory maximum. By contrast, the total-energy deviation remains $1.0\times10^{-4}\,\epsilon$ per atom even though the kinetic and potential deviations each reach about $7\times10^{-3}\,\epsilon$ per atom. Their near cancellation identifies the dominant error as a phase mismatch in the kinetic--potential exchange rather than energy loss. The model drift remains within a factor of twenty of the reference drift (Table~\ref{tab:energy_errors}). The main limitation at $T=0.20$ is therefore localized late-time derivative error, not a breakdown of conservation. This is the failure mode targeted by the late-stage top-$k$ and tail residual emphasis of Section~\ref{sec:methods_topk_tail} and discussed further in Section~\ref{sec:discussion}.

\subsection{Scaling to a denser 500-atom system}\label{sec:n500_results}

The same physics-only framework, also using the bounded warp-ballistic ansatz of Eq.~\ref{eq:ansatz_bounded} with $\tau_{\mathrm w}=0.12$, was then evaluated on a substantially larger and denser system: $N=500$ atoms at number density $\rho=0.50$ and horizon $T=0.12$. This setting is harder than the dilute base case because the mean inter-particle separation is smaller, local packing correlations are stronger, and the Lennard--Jones force field varies more rapidly. It is also the regime in which physics-informed MD solvers have previously been most challenged: a prior data-assisted physics-informed neural network (PND) has been reported to deviate from the reference radial distribution function on the $500$-atom argon case.\citep{Pham2024}

Increasing the system to $N=500$ does not produce the degradation seen at the longer horizon. The dense run is the most accurate of the three principal cases. Its maximum position error is $1.18\times10^{-5}\,\sigma$, its maximum velocity error is $2.93\times10^{-4}\,\sigma/\tau_{\mathrm{LJ}}$, and every energy deviation remains below $10^{-5}\,\epsilon$ per atom (Tables~\ref{tab:heatmap_errors} and~\ref{tab:energy_errors}). The model drift is within a factor of four of the reference drift. The small position error is spread broadly across atoms, whereas the largest velocity errors are more localized. The energy decomposition and the liquid-like RDF remain close to the reference (Figs.~\ref{fig:energy_all}c and~\ref{fig:rdf_all}c). These results indicate that, at the state points studied, the present bottleneck is temporal horizon and derivative accuracy rather than atom count alone.

\FloatBarrier

\subsection{Trajectory and energy error statistics}\label{sec:error_fields}

Tables~\ref{tab:heatmap_errors} and~\ref{tab:energy_errors} make the central contrast explicit. Extending the horizon from $0.12$ to $0.20$ produces the largest increase in position and velocity error, while scaling from $50$ to $500$ atoms at $T=0.12$ does not. In the extended run, the kinetic and potential deviations are nearly two orders of magnitude larger than the total-energy deviation, which supports the phase-error interpretation. The final column of Table~\ref{tab:heatmap_errors} (Atoms $>\tfrac{1}{2}\,$Max.) further shows that, for position and velocity separately, only two of the fifty extended-run atoms have peak errors above half the corresponding trajectory maximum, whereas the much smaller dense-run position error is broadly distributed ($232$ of the $500$ atoms). The corresponding space--time fields are shown in Appendix~\ref{app:error_heatmaps}.

\begin{table}[!htbp]
\centering
\caption{Full-trajectory position and velocity error norms for the three principal runs. Max., $P_{95}$, and Median are defined in Section~\ref{sec:evaluation_protocol}. The final column measures error localization: for each field it counts the atoms whose largest error norm over the saved frames exceeds one half of the Max.\ entry in the same row. A small count means the worst-case error is carried by a few atoms; a count comparable to $N$ means the error is broadly distributed across atoms.}
\label{tab:heatmap_errors}
\small
\begin{tabular}{llcccc}
\toprule
Run & Field & Max. & $P_{95}$ & Median & Atoms $>\tfrac{1}{2}\,$Max. \\
\midrule
\multirow{2}{*}{Base, $N=50$, $\rho=0.15$, $T=0.12$}
 & position & $7.48\times10^{-5}$ & $4.32\times10^{-5}$ & $5.7\times10^{-6}$ & $3/50$ \\
 & velocity & $1.98\times10^{-3}$ & $8.26\times10^{-4}$ & $1.0\times10^{-4}$ & $2/50$ \\
\midrule
\multirow{2}{*}{Extended, $N=50$, $\rho=0.15$, $T=0.20$}
 & position & $3.02\times10^{-3}$ & $4.74\times10^{-4}$ & $5.9\times10^{-5}$ & $2/50$ \\
 & velocity & $1.57\times10^{-1}$ & $1.42\times10^{-2}$ & $1.1\times10^{-3}$ & $2/50$ \\
\midrule
\multirow{2}{*}{Dense, $N=500$, $\rho=0.50$, $T=0.12$}
 & position & $1.18\times10^{-5}$ & $8.61\times10^{-6}$ & $4.2\times10^{-6}$ & $232/500$ \\
 & velocity & $2.93\times10^{-4}$ & $1.38\times10^{-4}$ & $4.1\times10^{-5}$ & $52/500$ \\
\bottomrule
\end{tabular}
\end{table}

\begin{table}[!htbp]
\centering
\caption{Maximum per-atom energy deviations and total-energy drifts for the three principal runs. The deviation columns are $\max_t\lvert X_\theta(t)-X_{\mathrm{MD}}(t)\rvert/N$ for $X\in\{K,U,E\}$. The drift columns are $\max_t\lvert E(t)-E(0)\rvert/N$, evaluated separately for the model and reference. Reference drift is the intrinsic velocity--Verlet drift. Reference energies are taken from LAMMPS's internal energy output.}
\label{tab:energy_errors}
\small
\begin{tabular}{lccccc}
\toprule
Run & $\max\lvert\Delta K\rvert$ & $\max\lvert\Delta U\rvert$ & $\max\lvert\Delta E\rvert$ & Model drift & Ref.\ drift \\
\midrule
Base, $N=50$, $T=0.12$ & $8.20\times10^{-5}$ & $9.15\times10^{-5}$ & $3.61\times10^{-5}$ & $3.51\times10^{-5}$ & $4.15\times10^{-6}$ \\
Extended, $N=50$, $T=0.20$ & $7.08\times10^{-3}$ & $7.14\times10^{-3}$ & $1.00\times10^{-4}$ & $9.70\times10^{-5}$ & $5.56\times10^{-6}$ \\
Dense, $N=500$, $T=0.12$ & $9.76\times10^{-6}$ & $5.42\times10^{-6}$ & $8.53\times10^{-6}$ & $8.48\times10^{-6}$ & $2.22\times10^{-6}$ \\
\bottomrule
\end{tabular}
\end{table}

\subsection{Robustness across initial conditions}\label{sec:robustness}

To check that the reported results are not an artifact of one favorable starting configuration, each system is regenerated from $N_{\mathrm{IC}}=5$ independent initial conditions. The initial conditions differ only in the random atom insertion and the initial Maxwell--Boltzmann velocities. Each is independently untangled, minimized, equilibrated, and given its own NVE reference, exactly as in Appendix~\ref{app:ic_grid}. A separate DINaMo network is trained for each initial condition, with all architecture, loss, and optimizer settings held fixed within each system class. All five base checkpoints were available. No checkpoint was found for the extended-horizon run \texttt{ic\_004} or for the dense run \texttt{ic\_002}, so those two aggregates contain four initial conditions each.

For each system, Table~\ref{tab:ic_robustness} reports the across-IC mean and sample standard deviation of the whole-trajectory position RMSE, the mean maximum total-energy error per atom, and trajectory containment at $0.01\,\sigma$. We report the sample standard deviation rather than a confidence interval because only four or five initial conditions are available and the aim is to expose sensitivity to the initial state rather than estimate a population parameter. An atom is counted as contained only when $\max_b e_i^{(r)}(t_b)\leq0.01\,\sigma$. With $B$ saved frames, the position RMSE is
\begin{equation}
\mathrm{RMSE}_{r}
=
\left[
\frac{1}{3NB}
\sum_{b=1}^{B}\sum_{i=1}^{N}
\left\|
\Delta\mathbf r_i(t_b)
\right\|_2^2
\right]^{1/2},
\label{eq:trajectory_rmse}
\end{equation}
where $\Delta\mathbf r_i(t_b)$ is the minimum-image displacement between the predicted and reference positions of atom $i$ at frame $t_b$ (the convention of Appendix~\ref{app:error_heatmaps}) and $B$ is the number of frames in the evaluation window. The $3NB$ normalization distinguishes this statistic from an average of the per-atom error norms $e^{(r)}_i$ of Eq.~\ref{eq:evaluation_errors}.

The base and dense classes are robust across the available initial conditions. All $250$ pooled base atoms and all $2000$ pooled dense atoms remain within $0.01\,\sigma$ throughout their evaluation windows. In the base class, $90.4\%$ ($226/250$) also remain within $10^{-3}\,\sigma$. The dense class is especially consistent, with $99.8\%$ ($1996/2000$) of atoms remaining within $10^{-4}\,\sigma$ and a mean whole-trajectory RMSE of $6.04\times10^{-6}\,\sigma$. The extended class shows the clearest sensitivity. It retains $98.0\%$ trajectory containment at $0.01\,\sigma$, and the four remaining atoms stay within $0.05\,\sigma$. Its RMSE standard deviation exceeds its mean because one difficult initial condition dominates the spread. Thus the longer-horizon degradation persists across initial conditions, but its severity depends strongly on the local starting configuration.

\begin{table}[!htbp]
\centering
\caption{Robustness across independent initial conditions. Values are the across-IC mean and sample standard deviation of whole-trajectory position RMSE, the mean maximum total-energy error per atom, and pooled trajectory containment at $0.01\,\sigma$ (five independent ICs per case). The energy column is the across-IC mean of $\max_t\lvert E_\theta(t)-E_{\mathrm{MD}}(t)\rvert/N$, using the same deviation convention as Table~\ref{tab:energy_errors}.}
\label{tab:ic_robustness}
\small
\setlength{\tabcolsep}{6pt}
\begin{tabular}{@{}l l c c@{}}
\toprule
Case & Position RMSE ($\sigma$) & Max.\ energy error & Containment \\
     & mean (s.d.) & ($\epsilon$/atom), mean & at $0.01\,\sigma$ \\
\midrule
$N=50$,\; $\rho=0.15$,\; $T=0.12$  & $1.58\times10^{-4}$ $(2.09\times10^{-4})$ & $4.47\times10^{-4}$ & 100\% (250/250) \\
$N=50$,\; $\rho=0.15$,\; $T=0.20$  & $6.62\times10^{-4}$ $(1.18\times10^{-3})$ & $9.65\times10^{-5}$ & 98.0\% (245/250) \\
$N=500$,\; $\rho=0.50$,\; $T=0.12$ & $6.04\times10^{-6}$ $(2.69\times10^{-6})$ & $8.40\times10^{-6}$ & 100\% (2500/2500) \\
\bottomrule
\end{tabular}
\end{table}

The difficult extended-horizon initial condition also makes the closest interparticle approach inside the reported window, reaching $0.87\,\sigma$ at $t\approx0.10$ compared with $0.95\,\sigma$ for the most benign member. As an ancillary diagnostic over the complete $929$-frame stored reference, including times beyond the reported evaluation window, $475$ frames contain a pair closer than $0.95\,\sigma$, compared with a single frame for the most benign member. Its maximum total-energy error is $1.84\times10^{-4}\,\epsilon$ per atom, nearly twice the class mean. Kinetic and potential errors grow in antiphase after the close approach while the total energy remains bounded, again indicating a conservative phase error. This association is descriptive rather than causal. Within each class, the architecture, loss settings, epoch budget, and causal configuration are fixed across initial conditions, so the observed spread reflects the starting configuration rather than per-run retuning.

\FloatBarrier

\subsection{Positioning relative to prior neural approaches}\label{sec:comparison}

The most informative comparison for a physics-only solver is not a single accuracy number alone but what each method is given to learn from, together with how it treats time, scale, and conservation. Table~\ref{tab:comparison} organizes representative neural approaches along these axes.

\begin{table}[!htbp]
\centering
\scriptsize
\resizebox{\linewidth}{!}{%
\begin{tabular}{p{2.4cm}p{3.2cm}p{2.3cm}p{2.4cm}p{2.6cm}}
\toprule
Approach (family) & Supervision (training signal) & Time representation & Conservation handling & Demonstrated \\
\midrule
Generative equilibrium samplers \citep{Olsson2026,Bonneau2026} & target-distribution samples / energies & static (equilibrium; no dynamics) & implicit & equilibrium ensembles, not dynamical trajectories \\
Trajectory / operator surrogates \citep{Schreiner2023,Park2026} & MD rollouts & discrete time-stepping (autoregressive) & not enforced & data-driven trajectory emulation \\
PINN-MD, data-assisted (PND) \citep{Razakh2021} & physics residuals $+$ MD data & continuous-time (PINN) & soft penalty terms & argon; RDF deviates on dense $500$-atom case \citep{Pham2024} \\
Physics-informed SNN (NP-SNN) \citep{Pham2024} & explicit $L_2$ match to simulator position; the general loss notation also lists velocity, acceleration, and energy & continuous-time (spiking / LIF) & soft penalty terms & argon $50$/$500$ (window $\approx0.02$); binary mixture $125$ ($100$ A $+$ $25$ B) \\
DINaMo (this work) & physics residuals only: Newton residual vs.\ analytic force on predicted positions; momentum/energy drift vs.\ initial state; no simulator targets & continuous-time global $\mathbf r_\theta(t)$, fixed MD grid & exact initial conditions $+$ residuals $+$ AL energy rate & argon $50$ ($\rho=0.15$) and $500$ ($\rho=0.50$), $T$ up to $0.20$; trajectory, energy, and RDF \\
\bottomrule
\end{tabular}%
}
\caption{Positioning of DINaMo relative to representative neural approaches for molecular dynamics. The columns compare the information condition (supervision), the time representation, and the handling of conservation laws, together with the demonstrated scope, rather than headline accuracy, which is not directly comparable across systems, units, and supervision regimes.}
\label{tab:comparison}
\end{table}

Generative samplers\citep{Olsson2026,Bonneau2026} and trajectory or operator surrogates\citep{Schreiner2023,Park2026} are trained directly on simulator output. The two existing physics-informed MD solvers are subtler. PND\citep{Razakh2021} writes the equations of motion and conservation laws into its loss, but trains against ground truth produced by its own bundled molecular-dynamics engine. The spiking solver NP-SNN\citep{Pham2024} is described as data-free, but its objective contains an explicit $L_2$ term comparing the predicted position with the simulator trajectory. Its general loss notation also lists simulator-based velocity, acceleration, and energy terms, although the accompanying description is not fully consistent about how these quantities are formed. We therefore base the supervision classification on the unambiguous position-matching term, which alone is sufficient to make the objective trajectory-supervised. DINaMo's training signal consists only of physics residuals: the Newton residual compares $m\mathbf a_\theta$ to the analytic Lennard--Jones force $\mathbf F(\mathbf r_\theta)$ evaluated on the predicted positions, and the conservation terms penalize drift of momentum and total energy relative to their values at $t=0$ (the residuals of Eqs.~\ref{eq:loss_momentum_energy}). No simulator position, velocity, force, or energy value ever appears in the objective. In this precise sense DINaMo is trajectory-unsupervised, physics-supervised rather than data-supervised, whereas prior physics-informed MD solvers retain a data-matching component.

Surrogate emulators advance the system one discrete step at a time and are evaluated by rollout (chaining many predicted steps and comparing the resulting trajectory with the reference).\citep{Schreiner2023,Park2026} Both NP-SNN and DINaMo are continuous in time, but they differ in window length: the argon trajectories shown in the published figures for NP-SNN span a short interval (on the order of $0.02$ in reduced units),\citep{Pham2024} whereas DINaMo trains continuous windows to $T=0.12$ and $0.20$. DINaMo explicitly defines both velocity and acceleration as derivatives of a single position field. The NP-SNN presentation describes both direct position--velocity outputs and derivative-based relations, so the two kinematic parameterizations are not directly equivalent. The direct horizon-extension experiment further makes the growth-limiting factor explicit rather than hiding it inside an integrator. All three physics-informed solvers reach $500$ atoms, the regime that most stresses structural fidelity, and it is precisely here that the data-assisted PND was reported to deviate from the reference $g(r)$.\citep{Pham2024} DINaMo's $500$-atom case is a strongly structured liquid, with $\rho=0.5$ and a first-neighbour $g(r)$ peak of approximately $3.5$ (Fig.~\ref{fig:rdf_all}c). It reproduces this structure from physics residuals alone, without the simulator $g(r)$ ever entering training.

Direct cross-method accuracy comparisons remain difficult: reported errors depend on the physical state, horizon, normalization, supervision regime, and metric definition, and several of these details are not matched across the available studies. We therefore use the prior work to position the information condition, system scale, and reported temporal extent of the methods, rather than to claim a controlled numerical advantage. The most meaningful absolute yardstick available is internal: across the three principal runs, the model total-energy drift lies within a factor of four to twenty of the intrinsic drift of the velocity--Verlet integrator that generated the reference (Table~\ref{tab:energy_errors}). The contribution of DINaMo is therefore not a claim of state-of-the-art accuracy, but the finding that meaningful molecular trajectories, energies, and structural observables can emerge from physics-only supervision without trajectory, force, or energy labels.

Finally, considering physical fidelity, PND and NP-SNN treat conservation through soft penalties.\citep{Razakh2021,Pham2024} By contrast, DINaMo combines exact initial-condition enforcement with physics-based consistency residuals and adaptive energy-rate control, yielding stable energy behavior and accurate structural observables without simulator supervision.

\section{Discussion}\label{sec:discussion}

The central result of these experiments is that minimizing the
governing physical residuals can recover more than a curve that merely
satisfies the training residuals: it can produce a molecular trajectory that agrees with an
independent MD reference across several observable levels at once.
The predicted positions remain close to the reference, the kinetic and
potential energies undergo the corresponding exchange, and the radial
distribution function preserves the local pair structure. These
agreements are informative because the subsequent reference trajectory
is never used as a training target. Energy conservation is imposed as a
physical constraint, but the reference kinetic, potential, and total
energy profiles are withheld, while $g(r)$ is not part of the objective
at all. Their recovery therefore tests whether the learned path represents
the underlying dynamics rather than merely satisfying one selected
summary statistic.

The three experiments also clarify how the approximation
deteriorates as the problem becomes more demanding. At the base horizon,
the trajectory, energy profile, and structure are all reproduced with
small errors. Extending the same formulation to $T=0.20$ preserves the
overall trajectory and structural statistics, but exposes a more
sensitive failure mode in the derivatives. The kinetic and potential
energy deviations grow largely in antiphase, so their errors cancel in
the total energy even while the velocity error becomes appreciable.
This distinction is important: bounded total energy is a necessary
indicator of physical consistency, but it is not by itself sufficient
to establish that the phase of the molecular motion is correct.
Derivative-sensitive trajectory metrics therefore provide information
that an energy-conservation measure alone would conceal. The dense
$500$-atom experiment complements this result. Its small position errors
and recovered liquid-like $g(r)$ show that the method is not relying
only on the weakly correlated structure of the dilute system, but can
also preserve the collective packing information generated by a more
strongly varying many-body force field.

These results also illustrate the trade-off introduced by the
global trajectory representation. A conventional integrator constructs
the solution through a sequence of local updates, whereas DINaMo
optimizes one differentiable function over the complete window. This
avoids step-to-step propagation of local integration error and ties position, velocity, and
acceleration to one common representation, but transfers the numerical
difficulty from time-stepping stability to global optimization. Errors
at late times can remain hidden by averages over easier portions of the
window, and small defects in the position field are amplified by
differentiation. The hard initial-condition ansatz, causal weighting,
late-window residual emphasis, and energy-rate constraint address
different parts of this optimization problem: they fix the local motion
known at $t=0$, preserve temporal ordering, expose localized late-time
residuals, and suppress the energy-changing component of the dynamical
error.

Two practical choices made during development further reflect
this optimization viewpoint. Randomly resampled collocation points gave
broad temporal coverage, but the locations of the largest late-time
residuals changed between epochs, making targeted improvements difficult
to retain. Fixing the collocation grid made those residual locations
persistent and allowed causal and tail-focused optimization to act on a
stable set of times. Likewise, adaptive equalization of the heterogeneous
loss terms could weaken the intended physical hierarchy by allowing
auxiliary conservation terms to compete with Newton's equation. Fixed
weights preserved Newton's equation as the primary residual while the
remaining terms acted as consistency constraints. The present
experiments establish the performance of this complete formulation.
Component-wise ablations that quantify the contribution of each mechanism
are a natural follow-up.

The present study addresses early-time molecular dynamics:
windows long enough to test trajectory, energy, and structural fidelity,
but too short for the long-time observables of production simulation. The
longest learned interval is $T=0.20\,\tau_{\mathrm{LJ}}$, evaluated at 401 points with
$\Delta t=5\times10^{-4}\,\tau_{\mathrm{LJ}}$. This is a nontrivial extension relative
to the continuous windows reported for comparable physics-informed
neural MD solvers, but it remains short relative to conventional
production trajectories and to the times required to estimate diffusion,
structural relaxation, transport coefficients, or rare-event kinetics.
The current solver is also instance-specific: a separate set of network
parameters is optimized for each equilibrated initial state that defines
a molecular initial-value problem, rather than one trained model being
reused across an ensemble of initial states. The results additionally
concern one analytic shifted-force Lennard--Jones interaction rather than
the diversity of interaction models used in practical simulation.

These limitations identify distinct extension problems. Longer
physical time could be reached by continuation across overlapping
windows, by transferring a converged solution to a subsequent interval,
or by composing separately optimized trajectory segments. Reuse across
initial states instead requires amortization, so that the
initial-condition encoder learns a family of initial-value problems
rather than participating in a separate optimization for each one.
Architectures or correction stages designed explicitly for derivative
accuracy could address the late-time velocity bottleneck, while
multicomponent systems and richer interaction potentials would test how
far the same physics-supervised formulation extends beyond the present
benchmark. The immediate significance of the current results is not that
explicit MD integration has already been superseded, but that the
solution of a molecular initial-value problem initialized from an equilibrated state can be
recovered over a nontrivial early-time window from the initial state and
governing physics alone, without learning from a precomputed reference
trajectory.

\section{Conclusions}\label{sec:conclusions}

We introduced DINaMo, a trajectory-unsupervised,
physics-informed neural solver that represents a molecular trajectory as
one continuous differentiable function of time. For a given equilibrated
initial state defining the molecular initial-value problem, the model is
optimized using Newton's equation, momentum and energy conservation, and
an analytic Lennard--Jones interaction, with no subsequent
simulator-generated position, velocity, force, or energy used as a
supervisory target.

Within each reported Lennard--Jones argon experiment, a single
predicted path reproduces complementary aspects of the reference
dynamics: coordinate trajectories, kinetic--potential energy exchange,
bounded total energy, and radial structure. The base experiment
establishes accurate short-window recovery. The $T=0.20$, 401-point
experiment shows that physical coherence can persist over a longer
window while identifying late-time derivative error as the dominant
limitation. The dense $500$-atom experiment shows that the approach
can retain liquid-like pair structure in a larger, more strongly
correlated system. Together, these observations provide evidence that
physics supervision can constrain a neural representation strongly
enough for physically meaningful molecular motion to emerge without
rollout labels.

The present method is an early-time,
instance-specific neural solver: it resolves one molecular initial-value problem,
initialized from an equilibrated state, over a learned interval of up to $T=0.20\,\tau_{\mathrm{LJ}}$,
rather than providing a reusable model across initial states or a
long-time alternative to conventional MD. Extending the physical time covered by
the learned trajectory, amortizing the solver across families of initial
states, improving late-time derivative accuracy, and treating more
general interaction models are therefore the next development steps.
Within its current scope, the study establishes the main feasibility
claim: a physically meaningful molecular trajectory can be recovered as
the solution of an initial-value problem from its initial state and
governing physics, rather than learned by regression onto a precomputed
simulator trajectory.

\section*{CRediT authorship contribution statement}
\textbf{Petros Triantafyllos:} Software development, Methodology, LAMMPS simulation setup and execution, Investigation, and Writing, original draft. \textbf{Panagiotis Krokidas:} LAMMPS simulation support and Writing, review and editing. \textbf{Christoforos Rekatsinas:} Conceptualization, Supervision, Project administration, and Writing, review and editing.

\section*{Declaration of competing interest}
The authors declare that they have no known competing financial interests or personal relationships that could have appeared to influence the work reported in this paper.

\section*{Funding}
This research did not receive any specific grant from funding agencies in the public, commercial, or not-for-profit sectors.

\section*{Declaration of generative AI and AI-assisted technologies}
During the preparation of this work the authors used generative AI tools to assist with the development of the accompanying software and with literature search. After using these tools, the authors reviewed, tested, and edited the resulting code and content as needed, and they take full responsibility for the content of the publication.

\section*{Data availability}
The LAMMPS input files and the resulting output files for all machine-learning simulations discussed in this work are publicly available through Figshare: \href{https://doi.org/10.6084/m9.figshare.32516052}{\texttt{10.6084/m9.figshare.32516052}}.

\section*{Code availability}
The source code used to train and evaluate DINaMo is available on GitHub at \url{https://github.com/insane-group/Differentiable-Newtonian-Molecular-Solver}.

\clearpage
\appendix
\numberwithin{equation}{section}
\numberwithin{figure}{section}
\numberwithin{table}{section}

\section{Run parameters}\label{app:run_configurations}

Table~\ref{tab:run_parameters} records the run-level physical, architectural, and optimization parameters used for the three principal experiments.

\begin{center}
\begin{minipage}{\textwidth}
\centering
\captionof{table}{Run-level parameters for the principal experiments. All quantities are in reduced Lennard--Jones units.}
\label{tab:run_parameters}
\scriptsize
\setlength{\tabcolsep}{4pt}
\renewcommand{\arraystretch}{1.12}
\resizebox{\textwidth}{!}{%
\begin{tabular}{p{4.0cm}p{4.4cm}p{4.4cm}p{4.4cm}}
\toprule
Parameter & Base trained horizon & Direct horizon extension & Dense scaling \\
\midrule
System & $N=50$, $\rho=0.15$ & \textit{same as base} & $N=500$, $\rho=0.50$ \\
Box length $L$ & $6.9336$ & \textit{same as base} & $10.0000$ \\
Reduced temperature $\Theta$ & $2.50487911765$ & \textit{same as base} & $0.70$ \\
Reduced LJ constants & $m=\epsilon=\sigma=1$ & \multicolumn{2}{c}{\textit{same as base}} \\
Pair interaction & LJ cutoff $r_c=2.937$, force-shifted & \multicolumn{2}{c}{\textit{same as base}} \\
Reference / collocation timestep & $\Delta t=5\times10^{-4}$ & \multicolumn{2}{c}{\textit{same as base}} \\
Training horizon $T$ & $0.12$ & $0.20$ & \textit{same as base} \\
NVE steps / grid points & $240$ / $241$ & $400$ / $401$ & \textit{same as base} \\
Collocation strategy & fixed MD grid; endpoints included; no jitter or scrambling & \multicolumn{2}{c}{\textit{same as base}} \\
Initial condition & post-NVT state & \multicolumn{2}{c}{\textit{same as base}} \\
Trajectory ansatz & $a_0$ hard-initial-condition ansatz & warp-ballistic ansatz, $\tau_{\mathrm w}=0.12$ & \textit{same as extended} \\
Network width / hidden layers & $256$ / $3$ & \multicolumn{2}{c}{\textit{same as base}} \\
Time / IC encoder layers & $1$ / $1$ & \multicolumn{2}{c}{\textit{same as base}} \\
Activation & ISRU, $\phi(x)=x/\sqrt{1+\alpha x^2}$, $\alpha=1$ & \multicolumn{2}{c}{\textit{same as base}} \\
Optimizer / learning rate & Adam / $10^{-3}$ & \multicolumn{2}{c}{\textit{same as base}} \\
Weight decay / LR floor & $0$ / $10^{-6}$ & \multicolumn{2}{c}{\textit{same as base}} \\
Residual weights $(w_N,\,w_P,\,w_E)$ & $1.0$ / $0.02$ / $0.02$; $w_E$ ramped in over the first $150$ epochs & \multicolumn{2}{c}{\textit{same as base}} \\
Per-atom energy scale $s_E$ & $0.1\,\epsilon$ per atom; energy-rate scale $s_E/T$ & \multicolumn{2}{c}{\textit{same as base}} \\
Epoch budget & $10{,}000$ & $20{,}000$ & \textit{same as extended} \\
Causal weighting & enabled; $\epsilon_c=0.1$, range $[0.005,0.1]$ & \multicolumn{2}{c}{\textit{same as base}} \\
Causal driver composition & mean-normalized Newton $0.97$, energy $0.02$, energy-rate $0.01$ & \multicolumn{2}{c}{\textit{same as base}} \\
Causal annealing & tail fraction $0.25$, ratio threshold $1.5$, decay/growth $0.92/1.01$ & \multicolumn{2}{c}{\textit{same as base}} \\
Energy-rate constraint & augmented Lagrangian; $\mu_0=0.005$, growth $1.03$, update every $400$ epochs; $\mu_{\max}=10^{5}$, violation clip $1.0$ & \multicolumn{2}{c}{\textit{same as base}} \\
AL warmup / warmup weight & $3000$ epochs / $0.02$ & $6000$ epochs / $0.02$ & \textit{same as extended} \\
Tail residual emphasis & fraction $0.20$, start $0.65$, ramp $0.10$, $(w_E,w_N)=(0.05,0.20)$ & \multicolumn{2}{c}{\textit{same as base}} \\
Atom-time top-$k$ mining & enabled; fraction $0.20$, min $16$, start $0.85$ & \multicolumn{2}{c}{\textit{same as base}} \\
Normalization settings & force normalization off; energy/power normalization on; deviation normalization fixed & \multicolumn{2}{c}{\textit{same as base}} \\
RDF settings & $80$ bins, $r_{\max}=2.937$ & \multicolumn{2}{c}{\textit{same as base}} \\
\bottomrule
\end{tabular}%
}
\end{minipage}
\end{center}

\section{Evolution of the formulation}\label{app:formulation_changes}

The present formulation is the result of difficulties encountered in earlier variants and of the design changes that overcame them. Early experiments used shorter windows, randomized time collocation, and adaptive relative balancing of the heterogeneous residual terms. These were reasonable starting points, but each raised a difficulty in this initial-value setting: randomized collocation made late-time residuals less persistent from epoch to epoch, so improvements there did not accumulate, and automatic balancing sometimes weakened the intended hierarchy between Newton's equation and the auxiliary conservation constraints. Both difficulties were overcome by the choices below.

The final formulation therefore keeps collocation on the MD time grid, treats Newton's equation as the primary residual, uses conservation laws as fixed physical regularizers, and enforces energy-rate consistency with an augmented Lagrangian. Causal annealing and late-stage top-$k$ residual emphasis then determine where optimization pressure is applied within the trajectory window. This combination gives a more stable route from the initial state and force law to a useful short-time trajectory.

\section{Reference initial conditions and time-grid matching}\label{app:ic_grid}

DINaMo and the LAMMPS reference are required to solve the \emph{same} molecular initial-value problem, so the construction of the initial state and the alignment of the evaluation grid are part of the experimental definition rather than incidental detail. All quantities below are in the reduced Lennard--Jones units defined in Section~\ref{sec:methods_reference}.

\paragraph{Shared initial condition.} For each system the reference is built as follows. Five independent initial conditions are generated with atom-insertion seeds $12345+1000k$ and velocity seeds $987654+1000k$ for $k=0,\dots,4$. Exactly $N$ atoms are inserted into a cubic periodic box of side $L=(N/\rho)^{1/3}$ subject to an overlap guard (minimum pair separation $0.80\,\sigma$). Because random insertion can leave close contacts, a guarded untangling stage then integrates the true Lennard--Jones dynamics with a capped per-step displacement (\texttt{nve/limit}, maximum step $0.005\,\sigma$) at a reduced timestep of $10^{-5}\,\tau_{\mathrm{LJ}}$ for $10^{4}$ steps, followed by a conjugate-gradient minimization under the same potential. Throughout, the interaction is the shifted-force \texttt{lj/smooth/linear} pair style with cutoff $r_c=2.937$, for which the energy and pair force vanish smoothly at the cutoff. Velocities are drawn from a Gaussian distribution at the target reduced temperature $\Theta$ with the net linear and angular momentum removed, and the system is equilibrated by NVT dynamics for $25\,\tau_{\mathrm{LJ}}$ at the production timestep $\Delta t=5\times10^{-4}\,\tau_{\mathrm{LJ}}$ (i.e.\ $5\times10^{4}$ steps). The post-NVT configuration and velocities are written to a single file (\texttt{ar\_ic\_lj.dump}, one record of $\mathbf r_0$ and $\mathbf v_0$ per atom), and the accompanying metadata record $N$, $L$, $\Delta t$, $r_c$, the force-shift convention, $m$, $\epsilon$, $\sigma$, and $\Theta$.

This post-NVT state is the single physical state from which both calculations originate. For the base run, the $a_0$ ansatz of Eq.~\ref{eq:hard_ic_ansatz} imposes the initial position, velocity, and acceleration by construction, with $\mathbf a_0=\mathbf F_{\mathrm{LJ}}(\mathbf r_0)/m$ evaluated using the same shifted-force law as LAMMPS. For the extended and dense runs, the warp-ballistic ansatz of Eq.~\ref{eq:ansatz_bounded} imposes the initial position and velocity exactly, while the initial acceleration is enforced through the Newton residual under the same analytic force law. No evaluation run re-equilibrates or resamples the initial state.

\paragraph{Time-grid matching.} After equilibration the timestep counter is reset, an exact pre-NVE energy is recorded at $t=0$, and the NVE rollout is produced by velocity-Verlet with a dump stride of one integration step, storing unwrapped coordinates and velocities. The reference frames therefore sit exactly at $t_b=b\,\Delta t$ for $b=0,1,2,\dots$ with $\Delta t=5\times10^{-4}\,\tau_{\mathrm{LJ}}$. The collocation grid of DINaMo (Section~\ref{sec:methods_collocation}, Eq.~\ref{eq:fixed_grid}) is placed on these same instants, so predictions and reference frames are compared one-to-one with no temporal interpolation. The default generation records $928$ NVE steps, i.e.\ a reference trajectory to $T=0.464\,\tau_{\mathrm{LJ}}$. Each experiment uses the leading prefix corresponding to its training horizon ($241$ frames for $T=0.12$ and $401$ frames for $T=0.20$). DINaMo does not advance the system with a timestep. It represents the trajectory as a continuous function of $t$ and is simply queried at the reference instants. ``Matching the timestep'' therefore means that the model's collocation grid coincides with the reference dump grid, not that the two share a numerical integrator.

\section{Heatmaps}\label{app:error_heatmaps}

\begin{figure}[p]
\centering
\includegraphics[width=\textwidth,height=0.85\textheight,keepaspectratio]{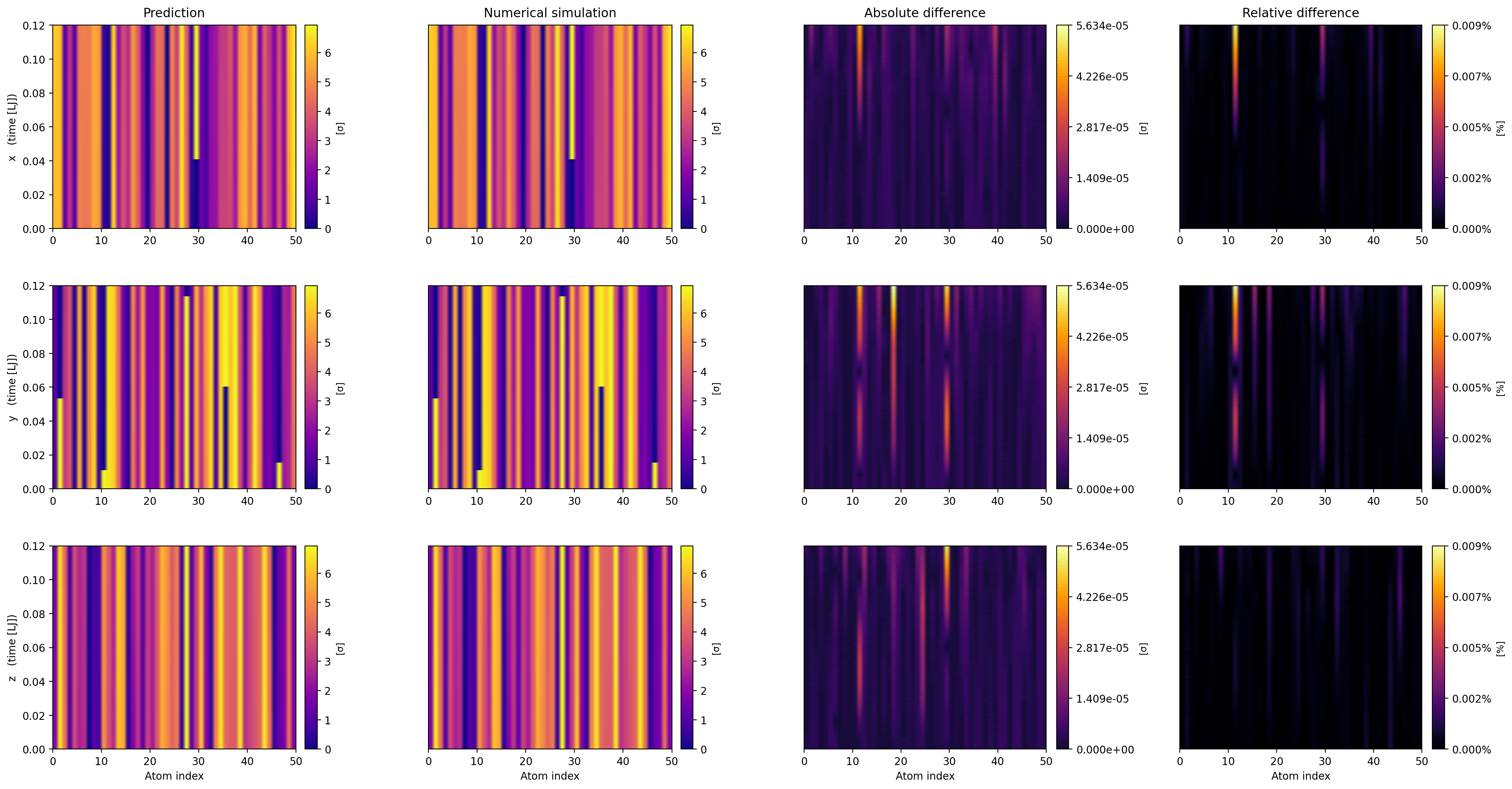}
\caption{Per-atom position-component heatmaps for the dilute base run ($N=50$, $\rho=0.15$, $T=0.12$).}
\label{fig:n50_pos}
\end{figure}

\begin{figure}[p]
\centering
\includegraphics[width=\textwidth,height=0.85\textheight,keepaspectratio]{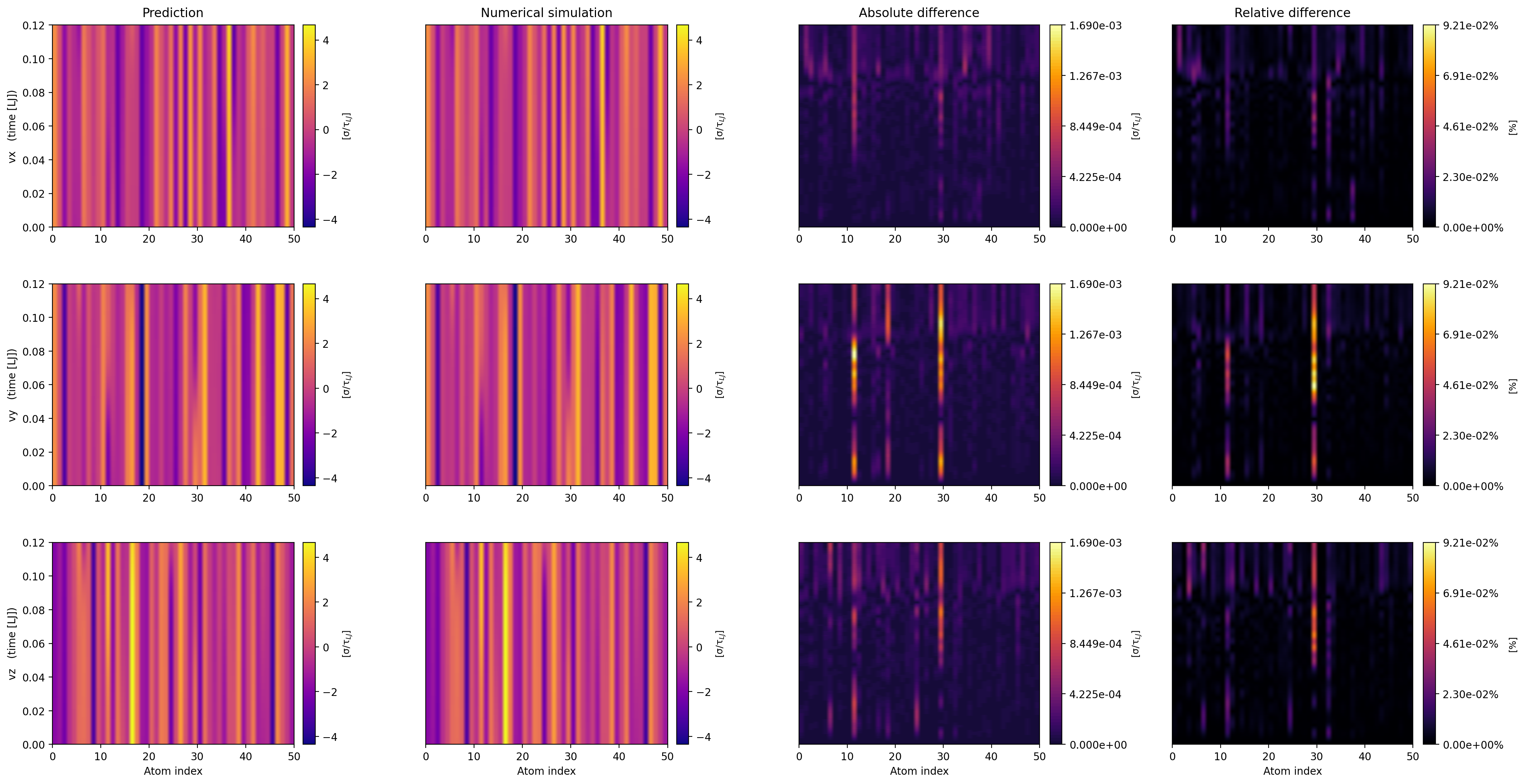}
\caption{Per-atom velocity-component heatmaps for the dilute base run ($N=50$, $\rho=0.15$, $T=0.12$).}
\label{fig:n50_vel}
\end{figure}

\begin{figure}[p]
\centering
\includegraphics[width=\textwidth,height=0.85\textheight,keepaspectratio]{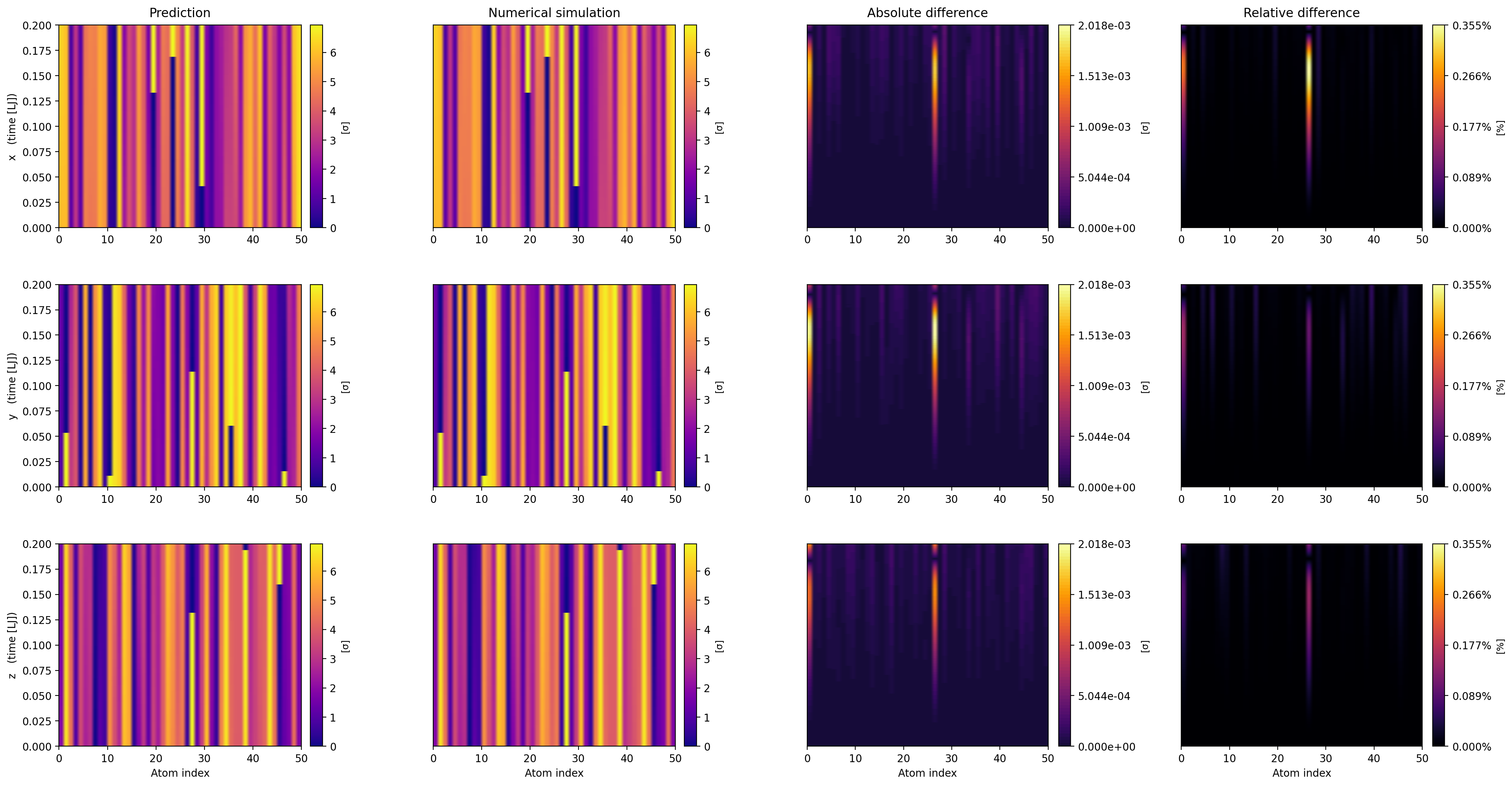}
\caption{Per-atom position-component heatmaps for the dilute extended-horizon run ($N=50$, $\rho=0.15$, $T=0.20$).}
\label{fig:n50_20_pos}
\end{figure}

\begin{figure}[p]
\centering
\includegraphics[width=\textwidth,height=0.85\textheight,keepaspectratio]{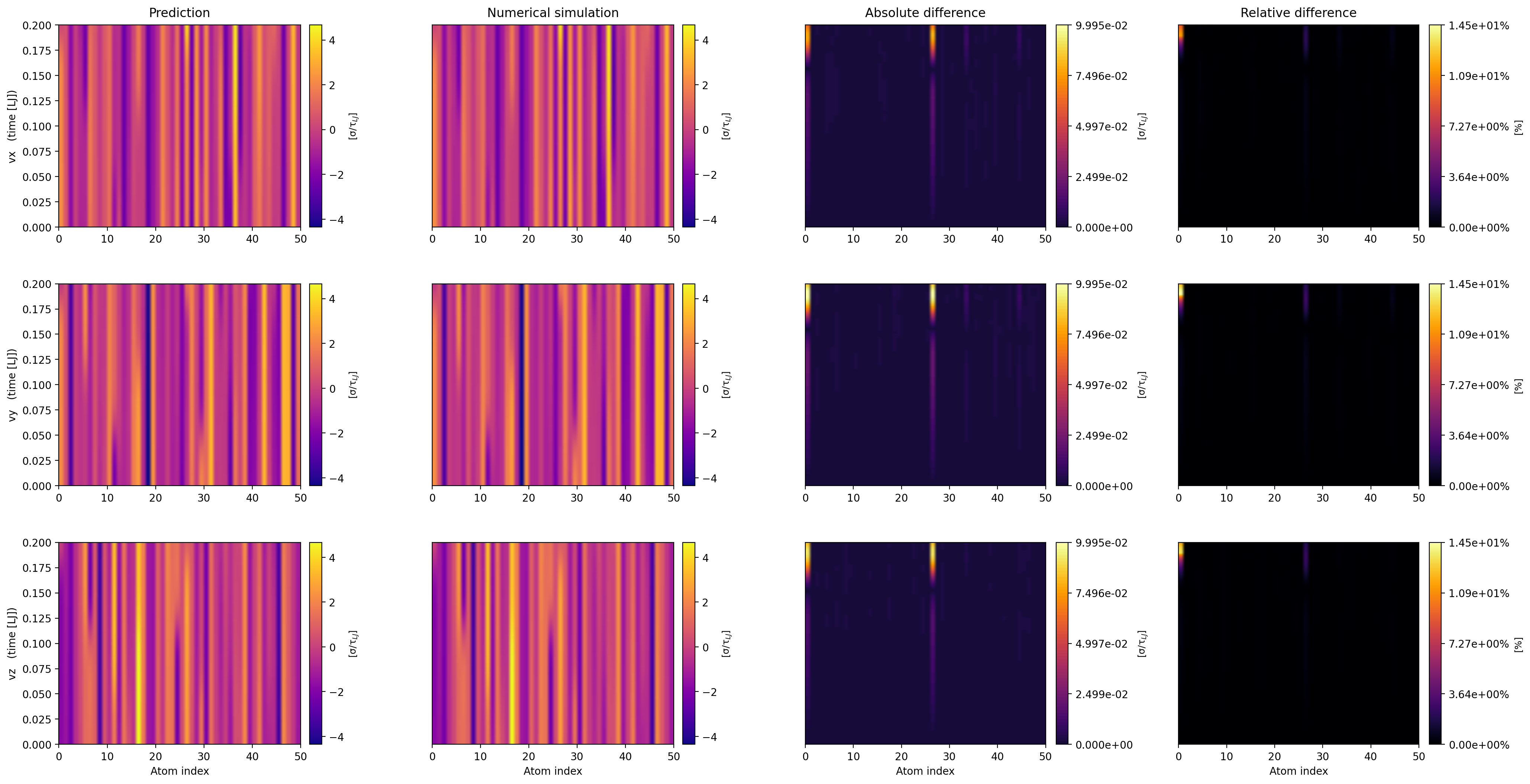}
\caption{Per-atom velocity-component heatmaps for the dilute extended-horizon run ($N=50$, $\rho=0.15$, $T=0.20$).}
\label{fig:n50_20_vel}
\end{figure}

\begin{figure}[p]
\centering
\includegraphics[width=\textwidth,height=0.85\textheight,keepaspectratio]{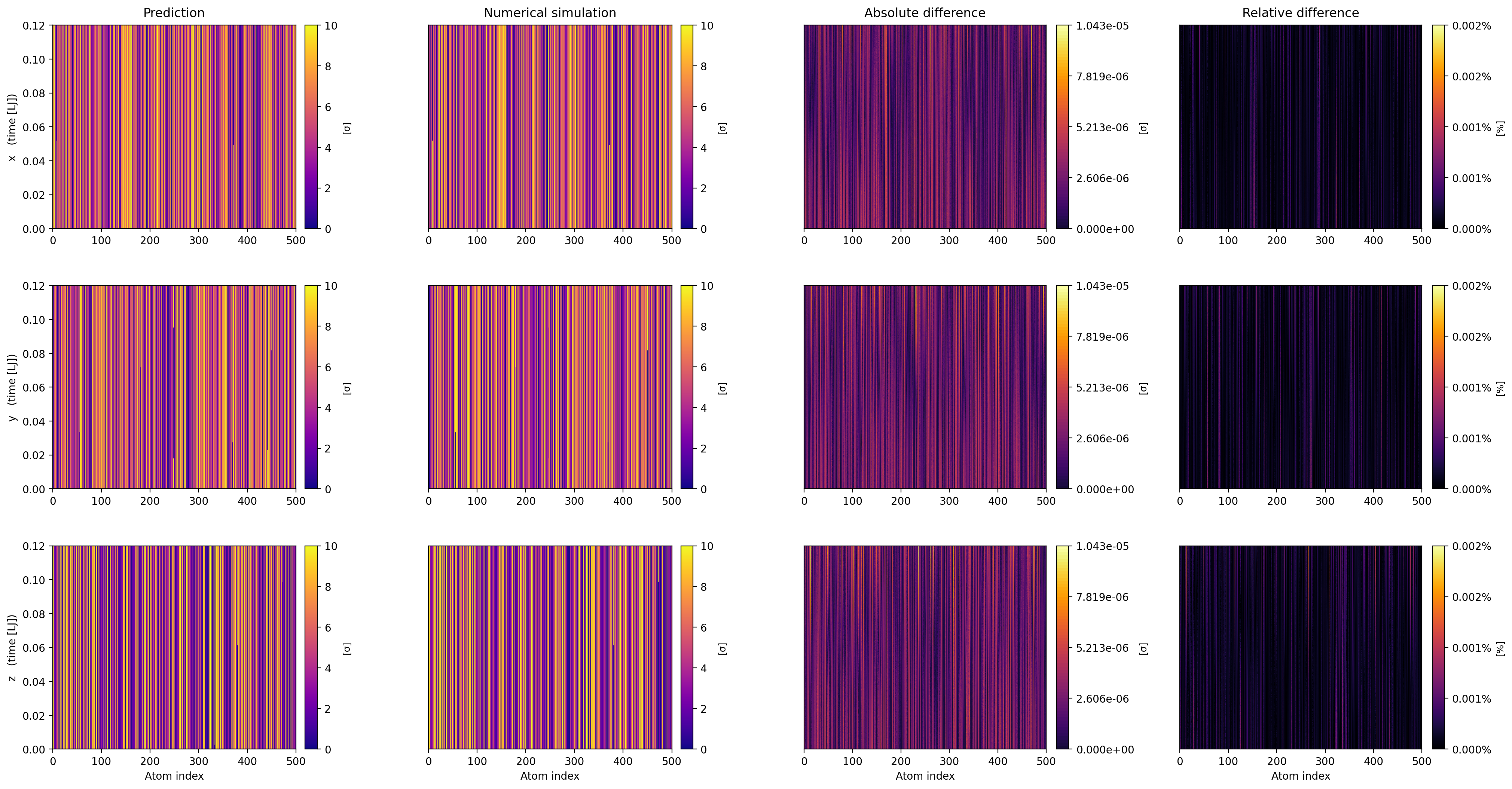}
\caption{Per-atom position-component heatmaps for the dense run ($N=500$, $\rho=0.50$, $T=0.12$).}
\label{fig:n500_pos}
\end{figure}

\begin{figure}[p]
\centering
\includegraphics[width=\textwidth,height=0.85\textheight,keepaspectratio]{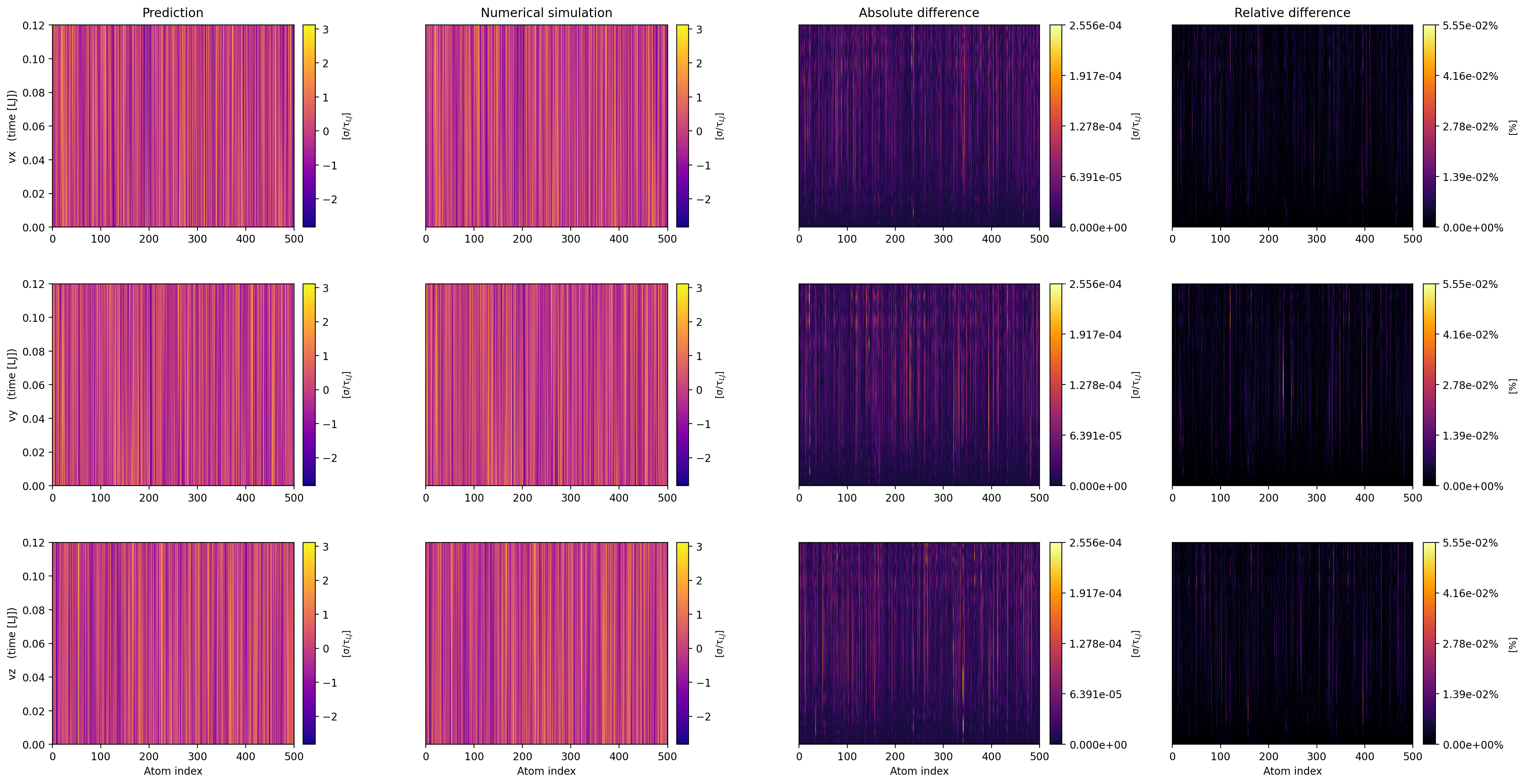}
\caption{Per-atom velocity-component heatmaps for the dense run ($N=500$, $\rho=0.50$, $T=0.12$).}
\label{fig:n500_vel}
\end{figure}

Figures~\ref{fig:n50_pos}--\ref{fig:n500_vel} show the per-atom componentwise errors for the three principal runs over their complete evaluation windows. The corresponding whole-trajectory statistics are summarised in Table~\ref{tab:heatmap_errors}. Position figures use rows $x$, $y$, and $z$, while velocity figures use rows $v_x$, $v_y$, and $v_z$. In every figure the columns show, from left to right, the DINaMo prediction, the LAMMPS reference, the absolute difference, and the floored relative difference. For positions, the absolute difference is computed with the minimum-image convention. The velocity heatmaps provide the more stringent comparison because velocities are obtained by differentiating the predicted positions.

Two conventions matter for reading these maps. First, position differences are computed with the minimum-image convention of the periodic box, $\Delta\mathbf r=\mathbf r_\theta-\mathbf r_{\mathrm{MD}}-L\,\mathrm{round}\!\big((\mathbf r_\theta-\mathbf r_{\mathrm{MD}})/L\big)$, so the reported difference is the true physical displacement and does not depend on the periodic image in which either trajectory happens to be expressed. The prediction and reference panels themselves display coordinates wrapped into $[0,L]$ for visual comparison. Second, the relative-difference panels show the per-axis absolute difference normalized by a floored reference magnitude and expressed in percent. For positions the reference is the magnitude of the box-centred coordinate. For velocities it is the reference speed of the atom. In both cases the reference is bounded below by a robust floor, defined as a fixed fraction of the corresponding root-mean-square scale, so that the ratio is not dominated by near-zero denominators at box seams or at velocity zero crossings. Dark regions therefore indicate small errors on both scales. Color ranges are set per figure from the data.

\FloatBarrier

\section{Conversion from reduced Lennard--Jones units to SI}\label{app:si_conversion}

The simulations and all reported errors are dimensionless. An SI interpretation is obtained only after choosing the Lennard--Jones length scale $\sigma$, energy scale $\epsilon$, and particle mass $m$ in SI units. The manuscript's reduced number density $\rho$ is the conventional $\rho^*$, and its reduced temperature $\Theta$ is the conventional $T^*$.

\begin{center}
\begin{minipage}{\textwidth}
\centering
\captionof{table}{Conversion of the reduced quantities used in this work to SI units. A starred or reduced numerical value is multiplied by the factor in the final column.}
\label{tab:si_conversion}
\small
\begin{tabular}{llll}
\toprule
Quantity & Reduced definition & SI unit & Conversion to SI \\
\midrule
Length & $r^*=r/\sigma$ & m & $r_{\mathrm{SI}}=r^*\sigma$ \\
Energy & $E^*=E/\epsilon$ & J & $E_{\mathrm{SI}}=E^*\epsilon$ \\
Mass & $m^*=m_{\mathrm{particle}}/m$ & kg & $m_{\mathrm{SI}}=m^*m$ \\
Time & $t^*=t/\tau_{\mathrm{LJ}}$ & s & $t_{\mathrm{SI}}=t^*\tau_{\mathrm{LJ}}$, $\tau_{\mathrm{LJ}}=\sigma\sqrt{m/\epsilon}$ \\
Number density & $\rho^*=n\sigma^3$ & m$^{-3}$ & $n_{\mathrm{SI}}=\rho^*/\sigma^3$ \\
Temperature & $T^*=k_B T/\epsilon$ & K & $T_{\mathrm{SI}}=T^*\epsilon/k_B$ \\
\bottomrule
\end{tabular}
\end{minipage}
\end{center}

For the common argon mapping $\sigma=3.405\times10^{-10}$~m, $\epsilon/k_B=119.8$~K, and $m=39.948$~u, the corresponding scales are $\epsilon=1.654\times10^{-21}$~J, $m=6.634\times10^{-26}$~kg, $\tau_{\mathrm{LJ}}=2.156\times10^{-12}$~s, and $\sigma^{-3}=2.533\times10^{28}$~m$^{-3}$. This mapping is illustrative: the reduced simulations themselves do not depend on a unique choice of SI parameterization. Under this mapping, $\Delta t=5\times10^{-4}$ corresponds to $1.08$~fs, the horizons $T=0.12$ and $0.20$ correspond to $0.259$ and $0.431$~ps, $\rho=0.15$ and $0.50$ correspond to number densities $3.80\times10^{27}$ and $1.27\times10^{28}$~m$^{-3}$, and $\Theta=2.504879$ and $0.70$ correspond to approximately $300$ and $83.9$~K, respectively.\citep{Frenkel2002}

\bibliographystyle{model1-num-names}
\begingroup
\sloppy
\emergencystretch=2em
\ifdefined\Urlmuskip\Urlmuskip=0mu plus 2mu\relax\fi
\bibliography{sn-bibliography}
\endgroup

\end{document}